\documentclass[twocolumn]{aastex61}
\pdfoutput=1 
\usepackage{amsmath,amstext}
\usepackage[T1]{fontenc}
\DeclareTextSymbol{\degre}{T1}{6}
\usepackage{apjfonts} 
\usepackage[figure,figure*]{hypcap}
\usepackage{subfigure}
\usepackage{gensymb}

\usepackage[normalem]{ulem}

\usepackage{color}

\usepackage{xspace}
\newcommand{\hst}{\textit{HST}\xspace}
\newcommand{\jwst}{\textit{JWST}\xspace}
\newcommand{\tablenotea}[1]{\parbox{6.6cm}{ \indent \footnotesize{\textsc{}~#1}}}

\newcommand\molhyd{H$\sb{2}$}
\newcommand\water{\ifmmode{{\rm H}\sb{2}{\rm O}}\else{H$\sb{2}$O}\fi}
\newcommand\methane{CH$\sb{4}$}
\newcommand\acetilen{C$\sb{2}$H$\sb{2}$}
\newcommand\carbdiox{CO$\sb{2}$}
\newcommand\pyratbay{\textsc{Pyrat Bay}}
\newcommand\taurex{\textsc{TauREx}}
\DeclareSymbolFont{UPM}{U}{eur}{m}{n}
\DeclareMathSymbol{\umu}{0}{UPM}{"16}

\shorttitle{WASP-43\MakeLowercase{\rm b} modelling}
\shortauthors{Venot et al.}

\begin{document}

\title{Global Chemistry and Thermal Structure Models for the Hot Jupiter WASP-43\MakeLowercase{b} and Predictions for \jwst}

\author[0000-0003-2854-765X]{Olivia Venot}
\affiliation{Laboratoire Interuniversitaire des Syst\`{e}mes Atmosph\'{e}riques (LISA), UMR CNRS 7583, Universit\'{e} Paris-Est-Cr\'eteil, Universit\'e de Paris, Institut Pierre Simon Laplace, Cr\'{e}teil, France}

\author{Vivien Parmentier}
\affiliation{AOPP, Department of Physics, University of Oxford}

\author{Jasmina Blecic}
\affiliation{NYU Abu Dhabi, Abu Dhabi, UAE}

\author{Patricio E. Cubillos}
\affiliation{Space Research Institute, Austrian Academy of Sciences, Schmiedlstr. 6, 8042, Graz, Austria}

\author{Ingo P. Waldmann}
\affiliation{University College London, Department of Physics and Astronomy, Gower Street, London WC1E 6BT, UK}

\author{Quentin Changeat}
\affiliation{University College London, Department of Physics and Astronomy, Gower Street, London WC1E 6BT, UK}

\author{Julianne I. Moses}
\affiliation{Space Science Institute, Boulder, CO, USA}

\author{Pascal Tremblin}
\affiliation{Maison de la Simulation, CEA, CNRS, Univ. Paris-Sud, UVSQ, Universit\'e Paris-Saclay, 91191 Gif-sur-Yvette, France}

\author{Nicolas Crouzet}
\affiliation{Science Support Office, Directorate of Science, European Space Research and
Technology Centre (ESA/ESTEC), Keplerlaan 1, 2201 AZ Noordwijk, The Netherlands}

\author{Peter Gao}
\affiliation{Astronomy Department, University of California, Berkeley, 501 Campbell Hall, MC 3411, Berkeley, CA 94720}

\author{Diana Powell}
\affiliation{Department of Astronomy and Astrophysics, University of California, Santa Cruz, CA 95064, USA}

\author{Pierre-Olivier Lagage}
\affiliation{AIM, CEA, CNRS, Universit\'e Paris-Saclay, Universit\'e Paris Diderot, Sorbonne Paris Cit\'e, UMR7158 F-91191 Gif-sur-Yvette, France}

\author{Ian Dobbs-Dixon}
\affiliation{NYU Abu Dhabi, Abu Dhabi, UAE}

\author{Maria E. Steinrueck}
\affiliation{Lunar and Planetary Laboratory, University of Arizona, 1629 E University Blvd, Tucson, AZ 85719, USA}

\author{Laura Kreidberg}
\affiliation{Harvard-Smithsonian Center for Astrophysics, 60 Garden Street, Cambridge, MA 02138}
\affiliation{Harvard Society of Fellows, 78 Mount Auburn Street, Cambridge, MA 02138}

\author{Natalie Batalha}
\affiliation{Department of Astronomy \& Astrophysics, University of California, Santa Cruz, CA 95064, USA}

\author{Jacob L. Bean}
\affiliation{Department of Astronomy \& Astrophysics, University of Chicago, 5640 S., Ellis Avenue, Chicago, IL 60637, USA}

\author{Kevin B. Stevenson}
\affiliation{Space Telescope Science Institute, 3700 San Martin Drive, Baltimore, MD 21218, USA}

\author{Sarah Casewell}
\affiliation{Dept. of Physics and Astronomy, Leicester Institute of Space and Earth Observation, University of Leicester, University Road,Leicester, LE1 7RH, UK}

\author{Ludmila Carone}
\affiliation{Max Planck Institute for Astronomy,  Koenigsstuhl 17, D-69117 Heidelberg, Germany}


\begin{abstract}
The \textit{James Webb Space Telescope} (\jwst) is expected to revolutionize the field of exoplanets. The broad wavelength coverage and the high sensitivity of its instruments will allow characterization of exoplanetary atmospheres with unprecedented precision. Following the Call for the Cycle 1 Early Release Science Program, the Transiting Exoplanet Community was awarded time to observe several targets, including WASP-43b. The atmosphere of this hot Jupiter has been intensively observed but still harbors some mysteries, especially concerning the day-night temperature gradient, the efficiency of the atmospheric circulation, and the presence of nightside clouds. We will constrain these properties by observing a full orbit of the planet and extracting its spectroscopic phase curve in the 5--12 $\mu$m range with \jwst/MIRI.
To prepare for these observations, we performed an extensive modeling work with various codes: radiative transfer, chemical kinetics, cloud microphysics, global circulation models, \jwst simulators, and spectral retrieval. Our \jwst simulations show that we should achieve a precision of 210 ppm per 0.1 $\mu$m spectral bin on average, which will allow us to measure the variations of the spectrum in longitude and measure the night-side emission spectrum for the first time. If the atmosphere of WASP-43b is clear, our observations will permit us to determine if its atmosphere has an equilibrium or disequilibrium chemical composition, providing eventually the first conclusive evidence of chemical quenching in a hot Jupiter atmosphere. If the atmosphere is cloudy, a careful retrieval analysis will allow us to identify the cloud composition.

\end{abstract}


\section{Introduction}

Giant planets that orbit very close to their host stars --- so-called ``hot Jupiters'' --- are expected to be tidally locked, with one hemisphere constantly facing the star, and one hemisphere in perpetual darkness.  The uneven stellar irradiation incident on such planets leads to strong and unusual radiative forcing, resulting in large temperature gradients and complicated atmospheric dynamics.  The atmospheric composition and cloud structure on these planets can, in turn, vary in three dimensions as the temperatures change across the globe, and as winds transport constituents from place to place.  Strong couplings and feedbacks between atmospheric chemistry, cloud formation, radiative transfer, energy transport, and atmospheric dynamics exist to further influence atmospheric properties.  The inherently non-uniform nature of these atmospheres complicates derivations of atmospheric properties from transit, eclipse, and phase curve observations.  Three-dimensional models that can track the relevant physics and chemistry on all scales --- both large and small distance scales, and large and small time scales --- are needed to accurately interpret hot Jupiter spectra.

Discovered in 2011 by \cite{hellier2011}, the hot Jupiter WASP-43b orbits a relatively cool K7V star (4\,520 $\pm$ 120 K, \citealt{gillon2012}\footnote{Note that the stellar effective temperature has been re-estimated to $T_{\mathrm{eff}}=4\,798~\pm$~216~K by \cite{sousa2018} after our models have been run.}). It has the smallest semi-major axis of all confirmed hot Jupiters and one of the shortest orbital periods (0.01526 AU and 19.5 h respectively, \citealt{gillon2012}). 
In addition to its exceptionally short orbit, WASP-43b is very good candidate for in-depth atmospheric characterization through transit thanks to its large planet-star radius ratio and the brightness of its host star, leading to a very good signal-to-noise ratio. The planet is also a good candidate for eclipse and phase curve observations thanks to the important flux ratio between the emission of the star and the exoplanet.

To date, many observations of the planet's atmosphere have been conducted from the ground \citep{gillon2012,wang2013,chen2014,murgas2014,zhou2014,jiang2016, hoyer2016} and from space \citep{blecic2014,kreidberg2014,stevenson2014,stevenson2017,ricci2015}. Most notably, orbital phase curves have been observed with both the Hubble Space Telescope (\hst) from $1.1$ to $1.7 \mu m$~\citep{stevenson2014} and with the \textit{Spitzer Space Telescope} at $3.6\mu m$ and $4.5\mu m$~\citep{stevenson2016}. Whereas these observations were able to constrain the dayside temperature structure and water abundances, they revealed the presence of a surprisingly dark nightside. Indeed neither Hubble nor Spitzer were able to measure the nightside flux from the planet. Poor energy redistribution~\citep{Komacek2016}, high metallicity~\citep{kataria2015}, disequilibrium chemistry~\citep{Mendonca2018chemistry} or the presence of clouds~\citep{kataria2015} were proposed to explain this mystery.

The \textit{James Webb Space Telescope} (\jwst) is expected to transform our understanding of the complexity of hot Jupiters atmospheres thanks to the numerous observations of transiting hot Jupiters it will perform. Rapidly after the launch and commissioning of the \jwst, exoplanet spectra will be obtained in the framework of the Transiting Exoplanet Community Early Release Science (ERS) Program \citep{bean2018}. Three hot Jupiters will be observed using the different instruments of the \jwst and the data will be available immediately to the community. Among these, WASP-43b is the nominal target that will be observed during the  sub-program ``MIRI Phase Curve''. A full orbit phase curve, covering two secondary eclipses and one transit, will be acquired with MIRI \citep{rieke2015}.
We will observe WASP-43b with MIRI during the Cycle 1 ERS Program developed by the Transiting Exoplanet Community \citep[PIs: N. Batalha, J. Bean, K. Stevenson;][]{stevenson2016, bean2018}. The MIRI phase curve is our best opportunity to probe the cooler nightside of the planet, determine the presence and composition of clouds, detect the signatures of disequilibrium chemistry and more precisely measure the atmospheric metallicity. The MIRI phase curve will be complemented by a NIRSpec phase curve (GTO program 1224, Pi: S. Birkmann). The later will provide a robust estimate of the water abundances on the dayside of the planet but will probably not be able to obtain decisive information about the nightside. MIRI will observe the planet at longer wavelengths where the thermal emission is more easily detectable and provide the first spectrum of the nightisde of a hot Jupiter. Correctly interpreting the incoming data will be, however, challenging. As shown by~\citet{Feng2016}, many pitfalls including biased detection of molecules can be expected if a thorough modelling framework has not been developed.

To prepare these observations, intensive work has been carried out by members of the Transiting Exoplanet ERS Team to model WASP-43b's atmosphere. In Section \ref{sec:strategy}, we present the methodology of this paper and explain how our different models interact with each other. In Section \ref{sec:models}, we present the models, parameters, and assumptions that have been used in this study. In Section \ref{sec:comp_models}, we show the results we obtained with these models concerning the thermal structure, the chemical composition, and the cloud coverage. In Section \ref{sec:JWST_simu}, we simulate the data that we expect to obtain with \jwst, and we perform a retrieval analysis in Section \ref{sec:retrieval}. Finally, the conclusions are presented in Section \ref{sec:conclusion}.

\section{Strategy}\label{sec:strategy}

\begin{figure}[!h]
 \label{fig:methodo}
 \includegraphics[width=\columnwidth]{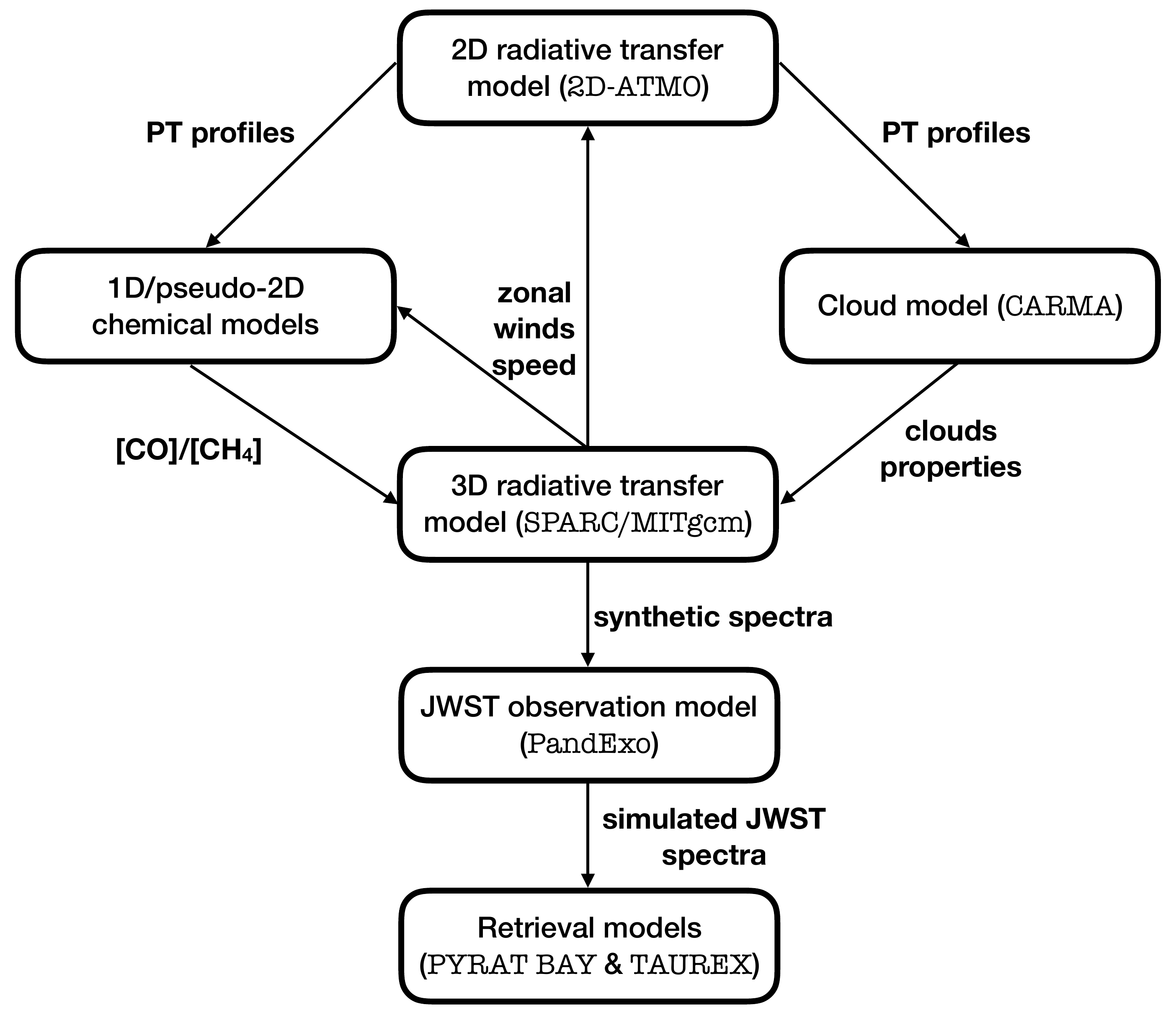}
 \caption{Strategy of modeling for this study representing the link between our various models.}
 \end{figure}

No single model is capable of simulating all the processes in an exoplanet atmosphere at once. The atmosphere must be modeled over many orders of magnitude in length scale (ranging from microphysical cloud formation to planet-wide atmospheric circulation). Planets are inherently 3D, and 1D models do not capture the expected variation in temperature, chemistry, or cloud coverage with location. It is not computationally feasible to include all the relevant effects in a single model. To simulate spectra for WASP-43b that are as complete as possible, we run a suite of models with a wide range of simplifying assumptions (1D, 2D, and 3D; equilibrium and disequilibrium chemistry). Each model is used as input for others. A major goal of this work is also to determine which modeling components are really necessary, or on the contrary can be substitute (i.e. we will see in Sect.\ref{sec:results_atmo_structure} that a 2D radiative/convective model can be a good alternative to a Global Circulation Model to calculate the thermal structure of a planet). Our methodology is represented on Figure \ref{fig:methodo} and detailed hereafter.

- Radiative/convective equilibrium models (2D) and general circulation models (GCM, 3D) have been used to compute the physical structure of the atmosphere. These two models indicate the temperatures all around the globe and thus quantify the day/night thermal gradient. The 3D model gives the zonal winds speeds that have been averaged (4.6 km.s$^{-1}$) and used by the 2D radiative/convective model as well as the pseudo-2D chemical model. Assuming a thermochemical equilibrium composition, both give very similar results (see Sect.~\ref{sec:results_atmo_structure}). Thus the 2D thermal profiles have been used as inputs in the chemical kinetics models. To simulate a disequilibrium composition, the 3D model uses the outputs of the chemical kinetics models, e.g. the average [CO]/[CH$_4$] ratio. For the cloudy modeling, the 3D model uses the findings of the microphysical cloud model, e.g. the nature of the clouds. Finally, the main outputs of the 3D model are the planetary spectra that have been binned by our \jwst simulator and then used to perform the retrieval. 

- Chemical kinetics models (1D and pseudo-2D) have been used to determine the chemical composition of the atmosphere, taking into account a detailed chemistry and out-of-equilibrium processes (photochemistry and mixing). These models predict whether the atmosphere is in thermochemical equilibrium. The comparison between the 1D and the pseudo-2D model enables an assessment of the influence of horizontal circulation on the chemical composition and whether we should expect a gradient of composition between the day and night sides. The inputs necessary for these models are the 2D thermal profiles calculated with the 2D radiative/convective model, as well as the zonal wind speed, whose average value comes from the ``equilibrium'' GCM model. In return, the GCM model uses the findings of the chemical kinetics model to set up the [CO]/[CH$_4$] in the ``quenched'' chemistry 3D model.

- Properties of clouds in the atmosphere of WASP-43b have been calculated using a cloud microphysical model. The findings of this model are informative about the possible location of clouds in the atmosphere, as well as the particle size and the nature of clouds. Determining the cloud coverage of the atmosphere is crucial as clouds tend to block the emitted flux of the planet and thus make harder the analyse of observational data. The cloud model uses as inputs the 2D thermal profiles. The outputs of this model (the nature of clouds and the particle size) was used in the cloudy 3D model.

-  Another series of models have been used to simulate the observations and their spectral analysis. The synthetic emission spectra (dayside and nightside) predicted by the GCM have been put through an instrument simulator to reproduce the instrumental noise and resolution of the \jwst instrument MIRI. The outputs of this simulator are then directly used by retrieval models.

- Finally, we used retrieval models to determine what information we will be able to extract from the new observations. This last step also permits to raise the difficulties that we will have to face and thus highlight which efforts/precautions will be required for the analyse of WASP-43b data. These models use the outputs of the \jwst simulator.

All the models used at each step are presented and detailed in the following subsections. It is important to note that the findings of each model are dependent on the intrinsic assumptions and design of each model. For instance, the retrieval models can only retrieve information that these models contain. It does, however, not mean that the presence of other molecules is thereby excluded. The physical properties of WASP-43b and its host star used in all the models are gathered in Table~\ref{tab:properties}.

\begin{table}[!h]
\caption{Properties of WASP-43 system.} \label{tab:properties}
\centering
\begin{tabular}{l@{\hspace{1.1cm}}r}
\hline \hline
Parameter   & Value$^a$  \\
\hline
Stellar Mass & 0.717 ($\pm$ 0.025) M$_{\sun}$\\
Stellar Radius & 0.667 ($\pm$ 0.01) R$_{\sun}$ \\
Effective Temperature & 4\,520 ($\pm$ 120 ) K \\
\hline
Planetary Mass & 2.034 ($\pm$ 0.052) M$_{J}$  \\
Planetary Radius & 1.036 ($\pm$ 0.019)  R$_{J}$ \\
Semi-major Axis & 0.01526 AU \\
 \hline
 \hline
 \\
\end{tabular}
\tablenotea{\textbf{Reference.} $^a$ \cite{gillon2012}.}
\end{table}

\section{Description of the models}\label{sec:models}

\subsection{Radiative Transfer Model}
\label{sec:RT_Model}

\texttt{ATMO} is a 1D/2D atmospheric model that solves the radiative/convective equilibrium with and without irradiation from an host star. It has been used for the study of brown dwarfs and directly imaged exoplanets \citep{tremblin:2015aa,tremblin:2016aa,tremblin:2017ab,leggett:2016aa,leggett:2017aa}, and also for the study of irradiated exoplanets \citep{drummond:2016aa,wakeford:2017aa,evans:2017aa}. The gas opacity is computed by using the correlated-k method \citep{lacis:1991aa,amundsen:2014aa,amundsen:2017aa} including the following species in this study: H$_2$-H$_2$, H$_2$-He, H$_2$O, CO, CO$_2$, CH$_4$, NH$_3$, K, Na, Li, Rb, Cs from the high temperature ExoMol \citep{tennyson:2012aa} and HITEMP \citep{rothman:2010aa} line list databases. We use 32 frequency bins between 0.2 and 320 $\mu$m with 15 k-coefficients per bin. The chemistry is solved at equilibrium or out-of-equilibrium by a consistent coupling with the chemical kinetic network of \citet{venot:2012aa}. \texttt{1D-ATMO} has been recently benchmarked against \texttt{Exo-REM} and \texttt{petitCODE} \citep{baudino:2017aa}.

In this study, we have used \texttt{2D-ATMO}, an extension of \texttt{1D-ATMO} \citep{tremblin:2017aa} that takes into account the circulation induced by the irradiation from the host star at the equator of the planet. We have taken a Kurucz spectrum \citep{castelli2004} for WASP-43 with a radius of 0.667 R$_{\sun}$, an effective temperature of 4500 K, and a gravity of log($g$) = $4.5$. The magnitude of the zonal wind is imposed at the substellar point at 4 km/s and is computed accordingly to the momentum conservation law in the rest of the equatorial plane. The vertical mass flux is assumed to be proportional to the meridional mass flux with a proportionality constant $1/\alpha$; the wind is therefore purely longitudinal and meridional if $\alpha \rightarrow \infty$ or purely longitudinal and vertical for $\alpha \rightarrow 0$. As in \citet{tremblin:2017aa}, a relatively low value of $\alpha$ drives the vertical advection of entropy/potential temperature in the deep atmosphere that can produce a hot interior, which can explain the inflated radii of hot Jupiters. A high value of $\alpha$ will produce a "cold" deep interior as in the standard 1D models. In this study, we have used two values of $\alpha$, 10 and 10$^4$ to explore these two limits. The simulation with $\alpha$=10$^4$ should be more representative of WASP-43b since the planet is not highly inflated.

\subsection{3-D Circulation Models}
\label{sec: 3-D Models}

\texttt{SPARC/MITgcm} couples a state-of-the art non-grey, radiative transfer code with the \texttt{MITgcm}~\citep{showman2009}. The \texttt{MITgcm} solves the primitive equations on a cube-sphere grid~\citep{Adcroft2004}. It is coupled to the non-grey radiative transfer scheme based on the plane-parallel radiative transfer code of \cite{Marley1999}. The stellar irradiation incident on WASP-43b is computed with a Phoenix model \citep{hauschildt1999}. The opacities we use are described in \cite{Freedman2008}, including more recent updates \citep{Freedman2014}, and the molecular abundances are calculated assuming local chemical equilibrium~\citep{Visscher2010}. In the 3D simulation, the radiative transfer calculations are performed on 11 frequency bins ranging from 0.26 to 300~$\mu$m, with 8 k-coefficients per bin statistically representing the complex line-by-line opacities. For calculating the spectra the final \texttt{SPARC/MITgcm} thermal structure is post-processed with the same radiative transfer code but using a higher spectral resolution of 196 spectral bins~\citep{Fortney2006}. We initialize the code with the analytical planet-averaged pressure-temperature profile of~\citet{Parmentier2015}, run the simulation for 300 days and average all physical quantities over the last 100 days of simulation.  

Our baseline model is a solar-composition, cloudless model. We also performed simulations including the presence of radiatively active clouds and radiatively passive clouds following the method outlined in~\citet{Parmentier2016}. Finally, models assuming a constant [CO]/[CH$_4$] ratio were also performed, following the method outlined in \cite{steinrueck2018}. These latter models simulate out-of-equilibrium transport-induced quenching of CO and CH$_4$.

\subsection{Chemical Kinetics Models}\label{sec:chemicalmodels}

To address the variations of atmospheric chemical composition with altitude and longitude, we used both 1D and 2D chemical kinetics models. We describe here these two codes.

\subsubsection{1D chemical kinetics model}
The thermo-photochemical model developed by \cite{venot:2012aa} is a full 1D time-dependent model. This model takes into account a detailed chemical kinetics and the out-of-equilibrium processes of photodissociation and vertical mixing (eddy and molecular diffusion). The atmospheric composition is computed for a fixed thermal profile divided in discrete layers, solving the continuity equations for each species until steady-state is reached. No flux of species is imposed at the boundaries of the atmosphere. In this study, we used the C$_0$-C$_2$ chemical kinetic network, which contains $\sim$ 2000 reactions describing the kinetics of 105 species made of H, C, O, and N, with up to two carbon atoms. This chemical scheme has been developed in close collaboration with specialists in combustion and validated experimentally on wide ranges of pressure and temperature as a whole (i.e. not only each reaction individually) leading to a high reliability. Since we consider both direction (forward and reverse) for each reaction, in absence of out-of-equilibrium processes, thermochemical equilibrium is achieved kinetically. To these $\sim$2000 reactions, 55 photodissociations have been added to model the interaction of incoming UV flux with molecules. These reactions are of course not reversed due to the disequilibrium irreversible nature of this process. A complete description of the model and the chemical scheme can be found in \cite{venot:2012aa}. The kinetic model has been applied to several exoplanetary atmospheres \citep{venot:2012aa,venot2013,venot2014,venot2016,agundez2014ApJ,tsiaras2016,rocchetto2016} and the deep atmosphere of Saturn \citep{mousis2014}, Uranus \citep{cavalie2014,cavalie2017}, and Neptune \citep{cavalie2017}.

Using this 1D chemical kinetics model, we determined the chemical composition of the atmosphere of WASP-43b at different longitudes around the equator. The vertical temperature structure as a function of longitude is taken from the 2D radiative-transfer models (Section \ref{sec:RT_Model}) utilizing $\alpha$=10$^4$ since the planet is not highly inflated. We extended these profiles to higher altitudes using extrapolation, and the resulting profiles are shown in Fig.~\ref{fig:temp2D}. Each longitude has been computed until steady-state is reached, independently from each other, starting from the thermochemical equilibrium corresponding to the thermal structure. The stellar zenith angle varied with longitude. Given the characteristics of the host star, we estimated the incident stellar spectrum using the same Kurucz \citep{castelli2004} stellar model than in \texttt{2D-ATMO} for wavelengths $\ge$ 2200 \AA, an IUE spectrum of GL~15B scaled by a factor of 10 for the 1250-2200 \AA\ region, and the solar maximum spectrum of \cite{woods2002} scaled by a factor of 1.36 \citep[see][]{czesla2013} for wavelengths less than 1250 \AA.
Vertical transport operates through eddy and molecular diffusion, with an assumed eddy diffusion coefficient $K_{zz}$ profile that varies as $K_{zz}$ (cm$^2$ s$^{-1}$) = 10$^7 \, \left[ P ( \textrm{bar} ) \right]^{-0.65}$ throughout the bulk of the planet's infrared photosphere, independent of longitude, except $K_{zz}$ approaches a constant-with-altitude value of 10$^{10}$ cm$^2$ s$^{-1}$ at high altitudes and at pressures greater than 300 bar (Figure \ref{fig:eddy}). As a large uncertainty resides for this parameter, we chose to use an expression similar to the one adopted for HD 189733b based on GCM passive tracer transport \citep{agundez2014aa}. This method gives values for the vertical mixing much lower (up to 1000 times) than the classical root-mean-square (rms) method \citep{agundez2014aa,charnay2015}. A correct estimation of this parameter is crucial as it determines the quenching level and thus the molecular abundances of some fundamental species \citep[e.g.][]{miguel2014,venot2014,Tsai2017}. Solar proportions of elemental abundances are assumed \citep{lodders2010solar} with a depletion of 20\% in oxygen due to the sequestration in silicates and metals \citep{lodders2004}. No TiO/VO has been included.

\subsubsection{2D chemical kinetics model}
Based on the procedure of \cite{agundez2014aa}, for this study we developed a ``pseudo-2D'' chemical model to track how the atmospheric composition of WASP-43b would vary as a function of altitude \textit{and} longitude, and hence orbital phase. We assume that longitudes are not isolated from each other (as in the 1D chemical model) rather connected through a strong zonal jet. Tidal interactions between host stars and gas giant planets are expected to circularize the planet orbit and synchronize the rotation period of the planet \citep{Lubow97,GuillotShowman2002}. The timescale for this to happen is much shorter than the stellar lifetime when the planet orbit is shorter than 10 days such as for WASP-43b \citep[see Fig.1 of][]{Parmentier2015echo}. With one hemisphere constantly facing the star, the unequal stellar forcing produces strong zonal winds that transport heat and chemical constituents across the dayside and into the nightside of the planet, and back again (e.g., \citealt{showman2010}; see also the GCM results in Section \ref{sec:results_3D}). This zonal transport provides ``diurnal'' variation in temperatures and stellar irradiation from the point of view of a parcel of gas being transported by the winds.  If chemical equilibrium were to prevail in WASP-43b's atmosphere, the composition would be very different on the colder nightside in comparison to the hotter dayside. However, if the horizontal transport by the zonal winds is faster than the rate at which chemical constituents can be chemically converted to other constituents, then the composition can be ``quenched'' and remain more uniform with longitude (e.g., \citealt{cooper2006,agundez2014aa}). Our pseudo-2D model tracks the time variation in atmospheric composition as a function of altitude and longitude as an atmospheric column at low latitudes experiences this pseudo-rotation.

Our pseudo-2D thermo/photochemical kinetics model uses the Caltech/JPL \texttt{KINETICS} code \citep{allen1981} as its base, modified for exoplanets as described in \cite{moses2011,moses2013a,moses2013b,moses2016}, along with the pseudo-2D procedure described in \cite{agundez2014aa}. The model contains 1870 reactions (i.e., ~900 forward and reverse pairs) involving 130 different carbon-, oxygen-, nitrogen-, and hydrogen-bearing species whose rate coefficients have been reversed using the thermodynamic principle of reversibility \citep[see][]{visscher2011}. Photolysis reactions are included and have not been reversed. The reaction list is taken from the GJ~436b model of \cite{moses2013b}, and includes H, C, O, N species. Molecules with up to six carbon atoms are included, but the possible chemical production and loss pathways in the model become less complete the heavier the molecule. Note that the chemical reactions list used in the pseudo-2D model slightly differs from that of the 1D chemical model. The departures between these two chemical schemes and their implications on the calculated abundances have already been addressed in several studies \citep{venot:2012aa,moses2014,wang2016}: depending on the scheme used, the quench level and resulting quenched abundances of some species are somewhat different. For instance, for HD~209458b-like planets, the quenched mixing ratio of CH$_4$ will be twice as large with \cite{moses2013b}'s scheme as with \cite{venot:2012aa}'s. However, the goal of our study here is to qualitatively compare the expected longitudinal variation obtained with the 1D and pseudo-2D chemical models in order to evaluate the effect of horizontal circulation on the global composition and transport/eclipse observations, and to determine a rough CO/CH$_4$ ratio to be used in the 3D GCM. Comparing results between the 1D and pseudo-2D chemical models described here adequately addresses these goals. Other model inputs, including the boundary conditions and vertical diffusion coefficients adopted in the model, are identical to 1D chemical kinetics model described above.

\begin{figure}
\label{fig:eddy}
\includegraphics[width=\columnwidth]{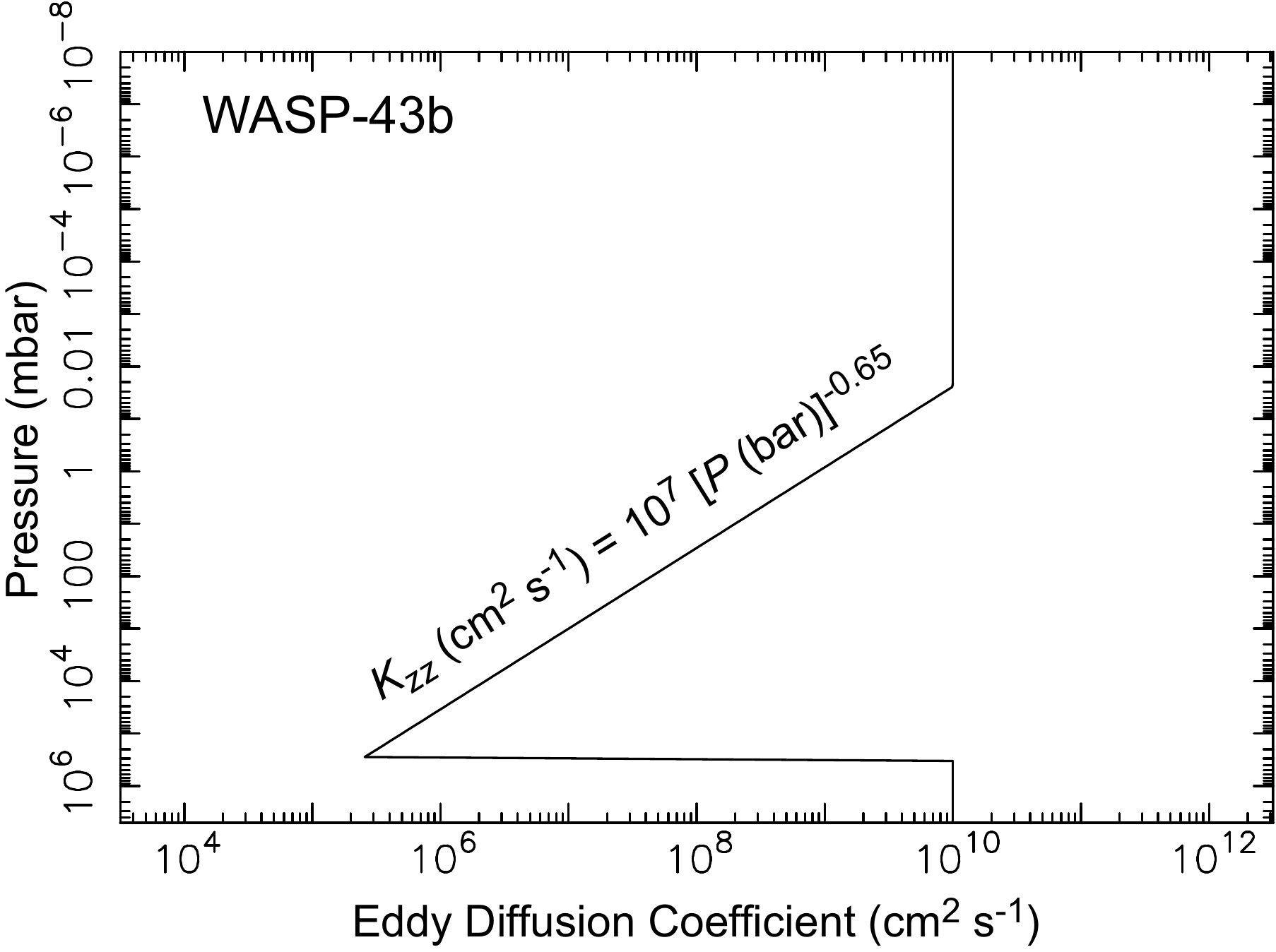}
\caption{The eddy diffusion coefficient profile adopted in our 1D and pseudo-2D thermo/photochemical models of WASP-43b. Our cloud microphysical model uses the same profile, except for pressures higher than 300 bar, where a lower limit value has been set at 10$^7$cm$^2$s$^{-1}$, to avoid numerical instabilities.}
\end{figure}

\begin{figure}
\label{fig:temp2D}
\includegraphics[width=\columnwidth]{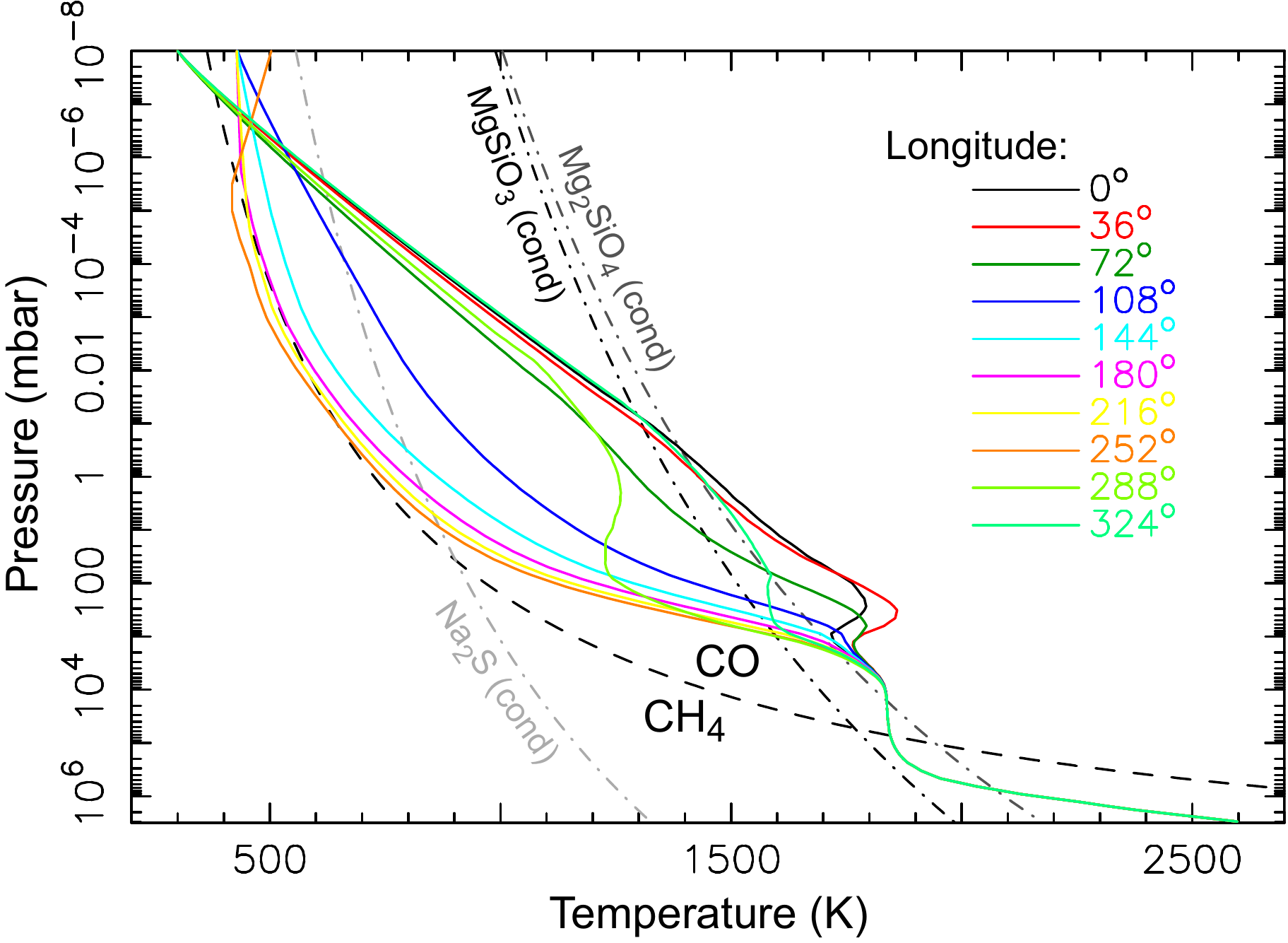}
\caption{Equatorial temperature profiles at different longitudes (colored solid lines, as labeled) adopted in our 1D and pseudo-2D thermo/photochemical models. Temperatures were derived from the 2D radiative-transfer model described in \cite{tremblin:2017aa}, for an assumed alpha value of 10,000 (uninflated radius, cold interior). The dashed black curve shows where CO and CH$_4$ would have equal abundance in chemical equilibrium for solar composition: CO dominates at lower pressures and higher temperatures, while CH$_4$ dominates at lower temperatures and higher pressures. The black and gray dot-dashed lines show the equilibrium condensation curves for MgSiO$_3$, Mg$_2$SiO$_4$, and Na$_2$S for a solar composition atmosphere.}
\end{figure}

Following \cite{agundez2014aa}, our pseudo-2D approach is to solve the 1D continuity equations for a vertical column of gas at the equator as it rotates around the planet with a constant average low-latitude jet speed of 4.6 km s$^{-1}$ (based on the GCM results described in Section \ref{sec:results_3D}). As it has been discussed in \cite{agundez2012,agundez2014aa}, assuming a uniform zonal wind is a good approximation for equatorial region, dominated by a superrotating jet \citep{kataria2015}. This approximation might be inadequate for latitudes towards the poles, when the circulation regime is more complex. Both the stellar zenith angle and atmospheric temperatures vary with time as the gas column rotates through different longitudes, both affecting the atmospheric chemistry. The vertical temperature structure as a function of longitude is fixed and is the same as that used in the 1D chemical kinetics model (Fig.~\ref{fig:temp2D}). The eddy diffusion profile ($K_{zz}$ coefficient profile) is also similar to that of the 1D chemical kinetics model (Fig. \ref{fig:eddy}). The planet is divided into different longitude regions, and the system of differential equations making up the continuity equations is integrated over the amount of time it would take a parcel of gas at the equator to be transported from one discretized longitude to the next discretized longitude. At that point, the mixing ratios as a function of pressure at the end of the first longitude calculation are fed in as initial conditions to the next longitude calculation, with its new thermal structure and incident UV flux that depends on the new zenith angle. The temporal evolution of this equatorial column of gas is followed for 20 full planetary ``rotations'' to provide sufficient time for the species produced photochemically at high altitudes to be transported down through the atmosphere to deeper regions where thermochemical equilibrium dominates. At that point, the ``daily'' longitude variations were consistent from one pseudo-rotation to the next. \cite{agundez2012} discuss the various advantages of beginning the pseudo-2D calculations at the hottest dayside conditions. Based on their discussion, we use the results of a 1D thermo/photochemical kinetics model for conditions at the substellar point (longitude 0\degre) that has been run long enough to reach steady state as our initial conditions for the pseudo-2D model.

\subsection{Cloud Microphysics Model}
\label{sec: Cloud Microphysics Models}

We use the Community Aerosol and Radiation Model for Atmospheres \citep[\texttt{CARMA};][]{turco1979,toon1988,jacobson1994,ackerman1995} to investigate the vertical and longitudinal distribution of clouds in the atmosphere of WASP-43b. While we do not include the resulting distributions in our GCM simulations, we will be able to extract insights into the effect of longitudinal temperature variations on cloud distributions. \texttt{CARMA} is a 1D cloud microphysics model that generates binned size distributions of aerosol particles as a function of altitude (pressure) in an atmospheric column by explicitly computing and balancing the rates of cloud particle nucleation, growth by condensation and coagulation, loss by evaporation, and transport by sedimentation, advection, and diffusion. This sets \texttt{CARMA} apart from simpler cloud condensation models \citep[e.g.][]{fegley1994,ackerman2001}, which assume cloud formation as soon as the condensate vapor saturates. The equations \texttt{CARMA} solves to evaluate the rates of these processes are presented in the Appendix of \citet{gao2018}. \texttt{CARMA} has been applied to aerosol processes across the solar system \citep{colaprete1999,barth2003,bardeen2008,gao2014,gao2017}, and has recently been used to simulate Al$_2$O$_3$, TiO$_2$, MgSiO$_3$ (enstatite), KCl, and ZnS clouds on exoplanets and brown dwarfs \citep{gao2018,powell2018,gao2018b}. 

For this work, we include additional condensates that have been hypothesized to dominate the condensate mass in exoplanet atmospheres, including Mg$_2$SiO$_4$ (forsterite), Fe, Cr, MnS, and Na$_2$S \citep{lodders1999,visscher2006,helling2008,Visscher2010}. Additional condensates are possible, as shown in grain chemistry models \citep[e.g.][]{helling2006}, but we do not treat them here, as it would be computationally prohibitive. In addition, to reduce the number of different condensates we assume that forsterite is the primary silicate condensate rather than modeling both forsterite and enstatite clouds. This is based on the argument that a rising parcel of vapor would see forsterite condense first due to it having higher condensation temperatures than enstatite; this depletes Mg and SiO, such that the enstatite cloud that forms above the forsterite cloud should have significantly lower mass. We discuss the implications of this assumption in the Sect. \ref{sec:results_clouds}. As with the treatment of enstatite in \citet{powell2018}, we assume forsterite and Fe clouds form by heterogeneously nucleating on TiO$_2$ seeds, as direct nucleation of these two species from vapor is slow \citep{helling2006,gao2018}. All other condensates are assumed to nucleate homogeneously.The saturation vapor pressures of Cr, MnS, and Na$_2$S are taken from \citet{morley2012}. The surface energy of Cr is calculated from the E\"otv\"os rule, while for MnS and Na$_2$S we assume the same surface energy as that of KCl. The size distribution for each condensate species is calculated separately, and so a distinct size distribution exists for each species.

As in the 1D and pseudo-2D chemical kinetics models, we use fixed pressure-temperature profiles described in Section \ref{sec:RT_Model} for our background atmosphere. All planetary parameters used are the same as those of the other models presented here to ensure consistency. In the cloud microphysics model, we use a very similar $K_{zz}$ profile as the one used in the chemical kinetics models (Figure \ref{fig:eddy}), except we set a minimum $K_{zz}$ of 10$^7$ cm$^2$ s$^{-1}$. This change only affects pressures $>$1 bar, where the chemical kinetics model $K_{zz}$ is $<$10$^7$ cm$^2$ s$^{-1}$. We also reduce the high $K_{zz}$ at pressures $>$300 bar to our minimum value. This was necessary to reduce model run time and numerical instabilities.  An atmospheric column is simulated at each longitude independently of each other, under the assumption that microphysical timescales are short compared to horizontal transport timescales, though this may not be the case for all pressure levels and particle sizes \citep{powell2018}. For each column, we investigate distinct clouds composed of Al$_2$O$_3$, TiO$_2$, Mg$_2$SiO$_4$, Fe, Cr, MnS, Na$_2$S, ZnS, and KCl, though which clouds actually form depends on which species is supersaturated and their nucleation rates. We assume solar abundances for the limiting elements of these clouds, which are, in the same order, Al, TiO$_2$, Mg, Fe, Cr, Mn, Na, Zn, and KCl. 

Importantly, the vertical, longitudinal, and particle size distributions computed by this model are not used to generate synthetic observations, as will be presented later in this work. This is due to the uncertainties in the material properties of some of the condensates (e.g. surface energies of MnS and Na$_2$S) and the way exoplanet clouds form, whether through homogeneous nucleation, heterogeneous nucleation on some foreign condensation nuclei, or grain chemistry \citep{helling2006}. Instead, results from \texttt{CARMA} will be helpful for informing general GCM and retrieval studies due to its ability to compute the relative abundances of different cloud species in the atmosphere of WASP-43b, thus indicating the species that affect the observations the most. Simplified cloud models can then be used to explore the parameter space around these results.
 
\subsection{JWST Observation Model}
\label{sec: JWST Observation Model}

WASP-43b is the primary target for the ``MIRI Phase Curve'' observation that will be carried out as part of the Transiting Exoplanet \jwst Early Release Science Program. The goal is to observe a full orbit of WASP-43b including two eclipses and one transit in the wavelength range 5--12$\; \mu$m at a resolution $R\sim100$ with MIRI LRS (Low Resolution Spectroscopy) in slitless mode \citep{kendrew2015}. In that program, the planetary emission spectra as a function of longitude will be measured and relevant atmospheric properties retrieved. We simulate the expected outcomes of this observation using the \texttt{PandExo}\footnote{\href{https://exoctk.stsci.edu/pandexo/}{https://exoctk.stsci.edu/pandexo/}} software program \citep{Batalha2017}. \texttt{PandExo} is a noise simulator specifically designed for transiting exoplanet observations with \jwst and \hst, and includes all observatory-supported time-series spectroscopy modes. 

The input parameters for the star and planet are those indicated in Table \ref{tab:properties}. The stellar spectrum is obtained from the \texttt{NextGen} \citep{hauschildt1999} grid interpolated at the $T_{\rm eff}$ and log($g$) of WASP-43 and is the same as used in the 3D \texttt{SPARC/MITgcm}. We consider a range of planetary emission spectra derived from the 3D \texttt{SPARC/MITgcm} model described in Section \ref{sec: 3-D Models}, with or without clouds, assuming thermochemical equilibrium or a quenched [CH$_4$]/[CO] ratio.
These simulations are performed with similar inputs as those used for the \jwst ERS Program proposal \citep[PIs: N. Batalha, J. Bean, K. Stevenson;][]{bean2018}: the radiative transfer models, the star and planet parameters and input spectra, and the observation parameters are the same (here we simulate a broader range of planetary spectra).
 
The planetary spectra are calculated from the emission integrated over the visible hemisphere. For this work, we simulate spectra with a spacing of 20\degre~in longitude and use them as inputs for \texttt{PandExo}. We consider that we observe each longitude during one eighteenth of the orbital period (1.08 hours) and we use a baseline of twice the eclipse duration because we will observe two eclipses (2.32 hours). In practice, the longitude 0\degre~will be in-eclipse so we may have to split the orbit slightly differently, but for these simulations we treat this longitude as the other ones.
The resolution and instrumental parameters are those of MIRI LRS. The wavelength range goes up to $\sim14 \; \mu$m but we consider only the 5--12$\; \mu$m range because the efficiency of LRS decreases significantly beyond 12$\; \mu$m. We use a saturation level of 80\% of the full well. The details of the noise modelling can be found in \citet{Batalha2017}.

\subsection{Retrieval models}\label{sec:retrieval_description}

To retrieve the atmospheric properties of WASP-43b, we use two models: {\taurex}\footnote{\href{https://github.com/ucl-exoplanets}{https://github.com/ucl-exoplanets}} \citep{waldmann2015tau,waldmann2015rex,rocchetto2016} and the Python Radiative Transfer in a Bayesian framework ({\pyratbay}\footnote{\href{http://pcubillos.github.io/pyratbay}{http://pcubillos.github.io/pyratbay}}, Cubillos et al. 2019, in prep., Blecic et al. 2019a,b, in prep.). Both, {\taurex} and {\pyratbay} are open-source retrieval frameworks that compute radiative-transfer spectra and fit planetary atmospheric models to a given set of observations. The atmospheric models consist of parameterized 1D profiles of the temperature and species abundances as a function of pressure, with atomic, molecular, collision-induced, Rayleigh, and cloud opacities. We decide to use two codes that don't use the same retrieval methods in order to compare the results obtained and raise the eventual biases that could emerge. We present hereafter the two codes.

\subsubsection{\taurex}
The {\taurex} model can retrieve equilibrium chemistry using the \texttt{ACE} code \citep{agundez2012} as well as perform so-called ``free'' retrievals where trace gas volume mixing ratios are left to vary as free parameters. For this study, all the retrieval models used the ``free chemistry'' method. The statistical sampling of the log-likelihood is performed using nested sampling \citep{Skilling2006,Feroz2009}.
{\taurex} is designed to operate with either absorption cross-sections or correlated-k coefficients. Both cross-sections and k-tables were computed from very high-resolution ($\mathrm{R}>10^6$). Cross-sections are calculated from ExoMol \citep{tennyson2016}, HITEMP \citep{rothman:2010aa} and HITRAN \citep{gordon2017} line lists using \texttt{ExoCross} \citep{yurchenko2018}. In particular, for this study we used the following elements: H$_2$O \citep{Barber2006}, CO \citep{rothman:2010aa}, CO$_2$ \citep{rothman:2010aa}, and CH$_4$ \citep{YurchenkoTennyson2014mnrasExomolCH4}, H$_2$ and He. Rayleigh scattering is computed for H$_2$, CO$_2$, CO and CH$_4$  \citep{bates1984,naus2000,bideau1973,sneepubachs2005} and collision induced absorption coefficients (H$_2$\,-\,H$_2$, H$_2$\,-\,He) are taken from \citet{richard2012}. Temperature and pressure dependent line-broadening was included, taking into account J-dependence where available \citep{Pine1992}. The absorption cross-sections were then binned to a constant resolution of $\mathrm{R}=15000$ and the emission forward models were calculated at this resolution before binning to the resolution of the data during retrievals.  {\taurex} can consider grey and Mie scattering clouds \citep{toon1981}, as well as the Mie opacity retrieval proposed by \citet{Lee2013}. The temperature-pressure profiles used in this study are parameterised by analytical 2-stream approximations \citep{Parmentier2014,Parmentier2015}.

\subsubsection{\pyratbay}
{\pyratbay} explores the parameter space via a Differential-evolution MCMC sampler \citep{CubillosEtal2017apjRednoise}, allowing both ``free'' and ``self-consistent'' (equilibrium chemistry) retrieval. 

The ``free'' retrieval fits for the thermal structure using the parameterized temperature profiles of~\citet{Parmentier2014} used by \citet[][]{LineEtal2013apjRetrievalI}, constant-with-altitude abundances for {\water}, {\methane}, and CO; and either one of the cloud parametrization models (detailed later in this Section).
In this study, we neglect {\carbdiox} because it does not contribute significantly in the spectrum of WASP-43b modelled by our Global Circulation Model on which the retrieval is performed, contrary to models of~\citep{Mendonca2018chemistry} where {\carbdiox} is proposed as a potential absorber on the nightside of the planet.

For ``self-consistent'' retrievals, we fit for the temperature and cloud parameters while assuming chemical equilibrium and solar elemental abundances. The chemical equilibrium is calculated with a newly developed open-source analytic thermochemical equilibrium scheme called \texttt{RATE}, Reliable Analytic Thermochemical-equilibrium Abundances \citep{CubillosBlecicDobbs-Dixon2019-RATE}, a similar, but more widely applicable approach than \citet{HengTsai2016}. For this study, we include only H$_2$O, CH$_4$, CO, CO$_2$, and C$_2$H$_2$ abundances and fix the elemental abundances to the solar ones of \citet{AsplundEtal2009}~\footnote{In general \texttt{RATE} is able to calculate the abundances of twelve atmospheric species (H$_2$O, CO, CO$_2$, CH$_4$, C$_2$H$_2$, C$_2$H$_4$, H$_2$, H, He, HCN, NH$_3$, and N$_2$) for arbitrary values of temperatures (200 to $\sim$2000 K), pressures (10$^{-8}$ to 10$^{3}$ bar), and C, N, O abundances (10$^{-3}$ to $\sim$10$^{2}$ $\times$ solar elemental abundances).}.

For the opacities {\pyratbay}  considers line-by-line opacities sampled to a constant wavenumber sampling of 0.3 cm$^{-1}$ for the four main spectroscopically active species expected at the probed wavelengths: {\water} \citep{rothman:2010aa}, {\methane} \citep{YurchenkoTennyson2014mnrasExomolCH4}, CO \citep{LiEtal2015apjsCOlineList}, and {\carbdiox} \citep{rothman:2010aa}. (The same species considered in {\taurex} but with different references for CO and H$_2$O). Since these databases consist of billions of line transitions, we first apply our repacking algorithm \citep{Cubillos2017apjCompress} to extract only the strong line transitions that dominate the opacity spectrum between 300 and 3000 K. Our final line list contains 5.5 million transitions. Additionally, {\pyratbay} considers Rayleigh opacities from {\molhyd} \citep{LecavelierDesEtangsEtal2008aaRayleighHD189}, collision-induced absorption from {\molhyd}--{\molhyd} \citep{BorysowEtal2001jqsrtH2H2highT, Borysow2002jqsrtH2H2lowT} and {\molhyd}--He \citep{BorysowEtal1988apjH2HeRT, borysowfrommhold1989b, borysowfrommhold1989a}. {\pyratbay} implements several cloud parameterization models: a simple opaque gray cloud deck at a given pressure, a thermal-stability cloud approach described in Blecic et al. (2019a, in prep.), and a kinetic, microphysical cloud parameterization model (Blecic et al. 2019b, in prep.). In all complex cloud models the cloud opacity is calculated using Mie-scattering theory \citep{toon1981}.

For cloud-free retrieval, {\pyratbay} uses a top pressure of a gray cloud deck in his cloud-free model. For cloudy retrievals in this study, we use our Thermal Stability Cloud (\texttt{TSC}) model (Blecic et al. 2019a, in prep.) to retrieve the longitudinal cloud structure \citep[see also][]{Kilpatrick2018}. The model is based on the methodology described in \cite{Benneke:2015} and \cite{ackerman2001} with additional flexibility in the location of the cloud base depending on the local metallicity of the gaseous condensate species just below the cloud deck.

The {\pyratbay} code explores the posterior parameter space with the Snooker differential-evolution MCMC \citep{terBraak2008SnookerDEMC}, obtaining between 1 and 4 million samples, with 21 parallel chains (discarding the initial 10\,000 iterations), while ensuring that the \citet{GelmanRubin1992} statistics remain at $\sim$1.01 or lower for each free parameter.

\section{Results of atmospheric models}\label{sec:comp_models}

\subsection{Atmospheric structure}\label{sec:results_atmo_structure}

 \begin{figure}[!h]
 \label{fig:3Deq_temps}
 \includegraphics[width=\columnwidth]{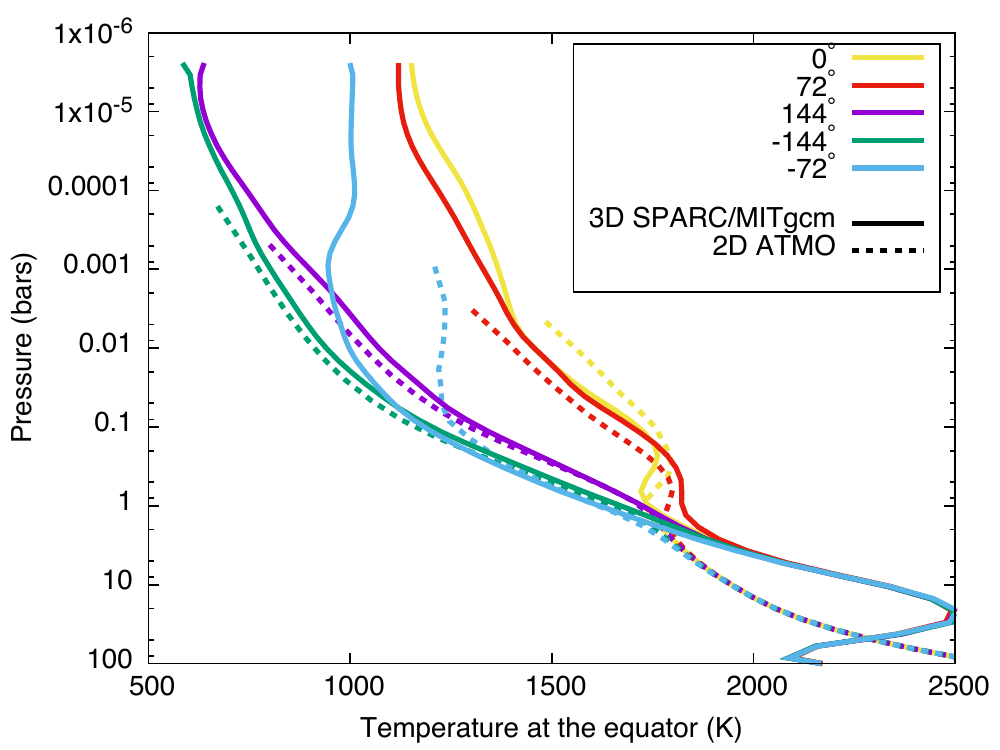}
  \includegraphics[width=\columnwidth]{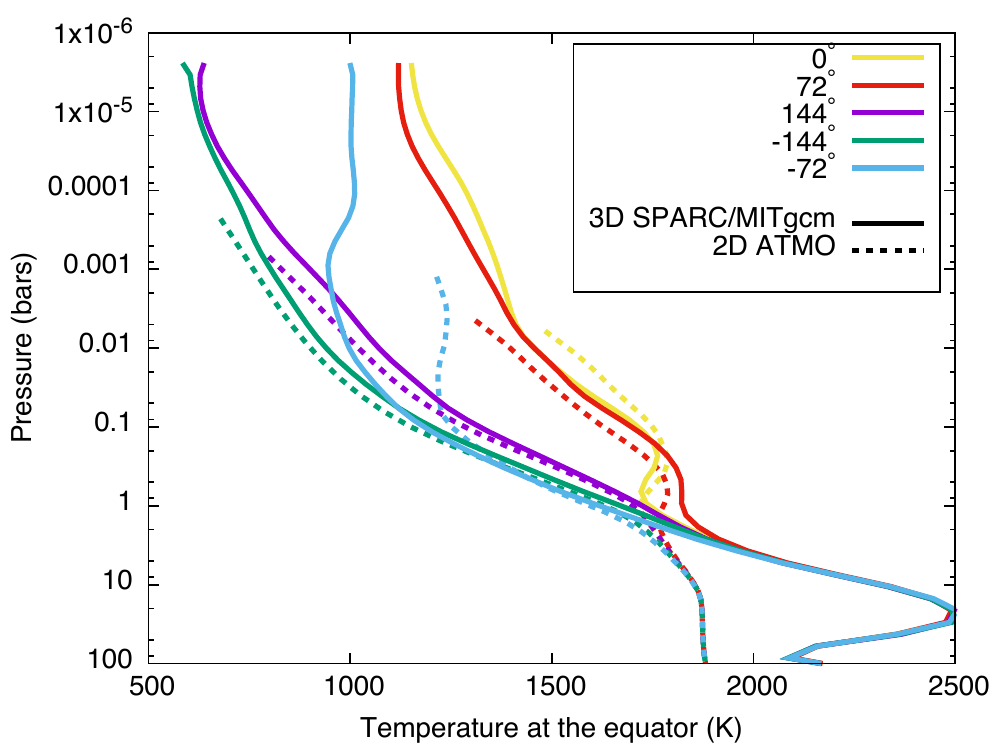}
 \caption{Comparison of the equatorial temperature structure predicted by the 3D \texttt{SPARC/MITgcm} model with the \texttt{2D-ATMO} model assuming $\alpha=10$ (top) and $\alpha=10^4$ (bottom).}
 \end{figure}
 
The pressure/temperature structure of the atmosphere can be constrained with 2D and 3D models. A comparison is given in Fig.~\ref{fig:3Deq_temps} between the 3D \texttt{SPARC/MITgcm} model and the \texttt{2D-ATMO} model with the two cases of interior: the hottest one ($\alpha$=10) and the coldest one ($\alpha$=10$^4$). For the two $\alpha$ values, the upper atmosphere is similar. In this part (pressures less than 1 bar), the agreement between the 2D and the 3D models is quite remarkable since there is no tuning of the 2D model on the 3D model apart from the choice of horizontal wind speed. Since the chemistry and radiative transfer models are independent between the two codes, this agreement is also a sign that convergence between different GCMs and 2D steady-state circulation models can be reached for the pressure/temperature structure at and above the photosphere. 
As explained in Sect. \ref{sec:RT_Model}, for pressures greater than 1 bar, the different $\alpha$ values lead to different temperatures in the deep atmosphere. Because the GCM is not fully converged at pressures larger than 10 bars, it likely produces spurious variations of the deep pressure-temperature profile. The shape of the deep flow structure and its influence in the upper atmospheric dynamics is still an active subject of research~\citep{mayne2014,Mayne2019,Carone2019} and out of the scope of this paper. Given that both the GCM and the 2D model produce very similar thermal structure in the observable atmosphere we decided to use the outputs of the 2D model as inputs for the chemical and cloud formation models. We opted for the cold interior model ($\alpha$=10$^4$) based on the ground that the planet does not appear to be highly inflated. However, a more thorough investigation of the deep thermal structure on the observable cloud properties~\citep[e.g.][]{powell2018} will be needed in the future to interpret the observations.

 
\subsection{Chemical composition}\label{sec:results_chemistry}

\begin{figure}[!h]
\label{fig:abun_chem}
\includegraphics[width=\columnwidth]{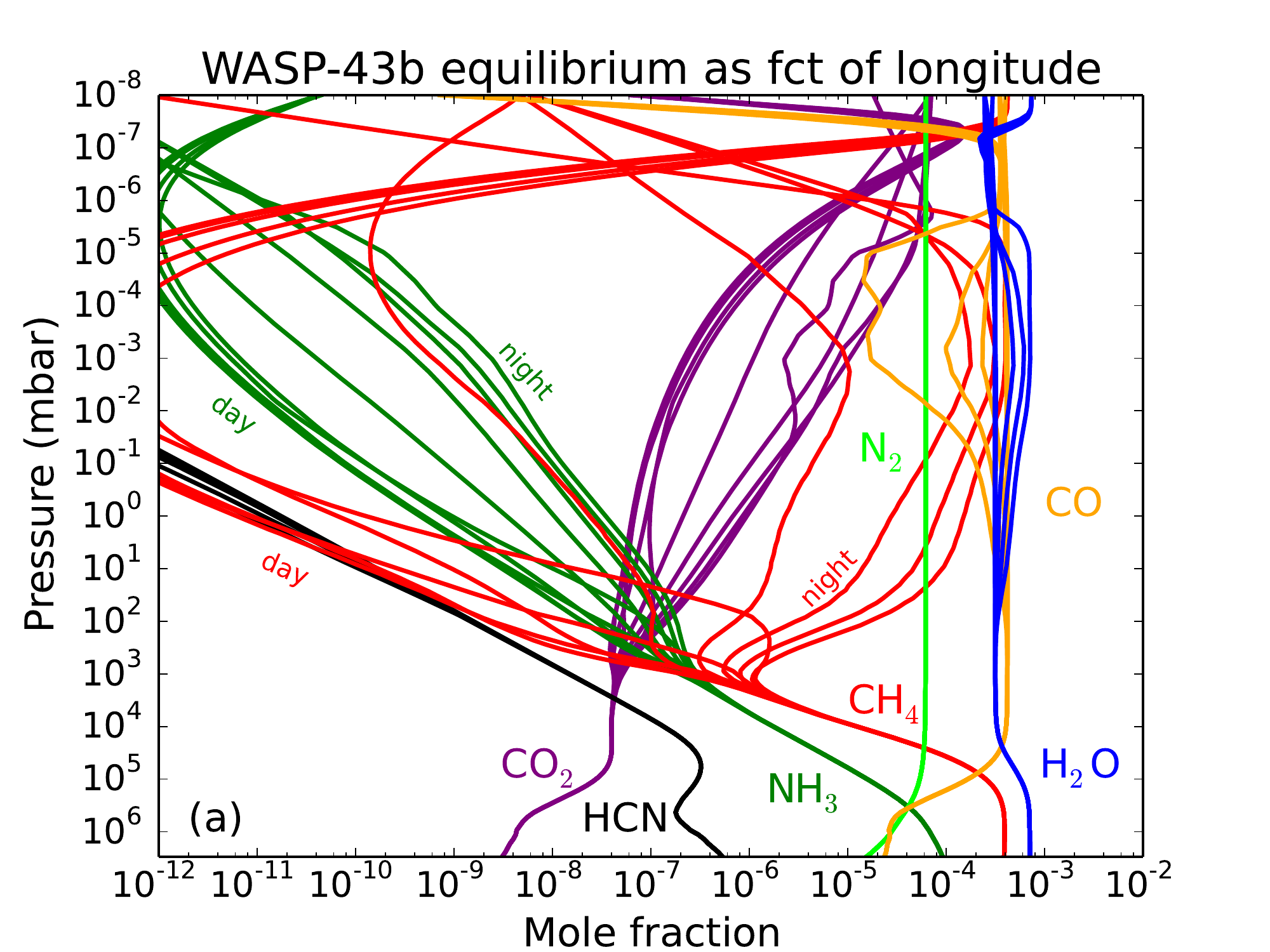}\\
\includegraphics[width=\columnwidth]{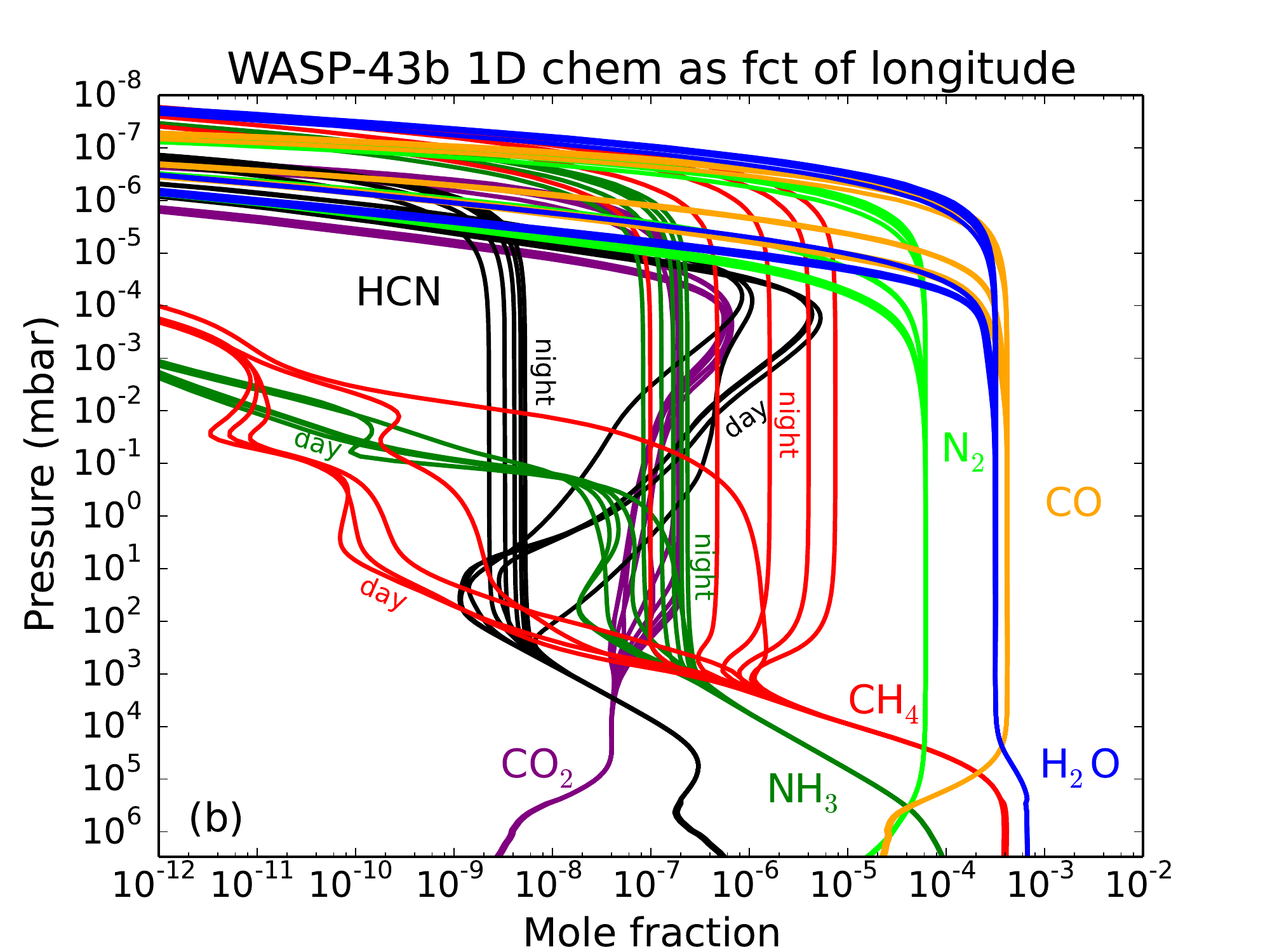}\\
\includegraphics[width=\columnwidth]{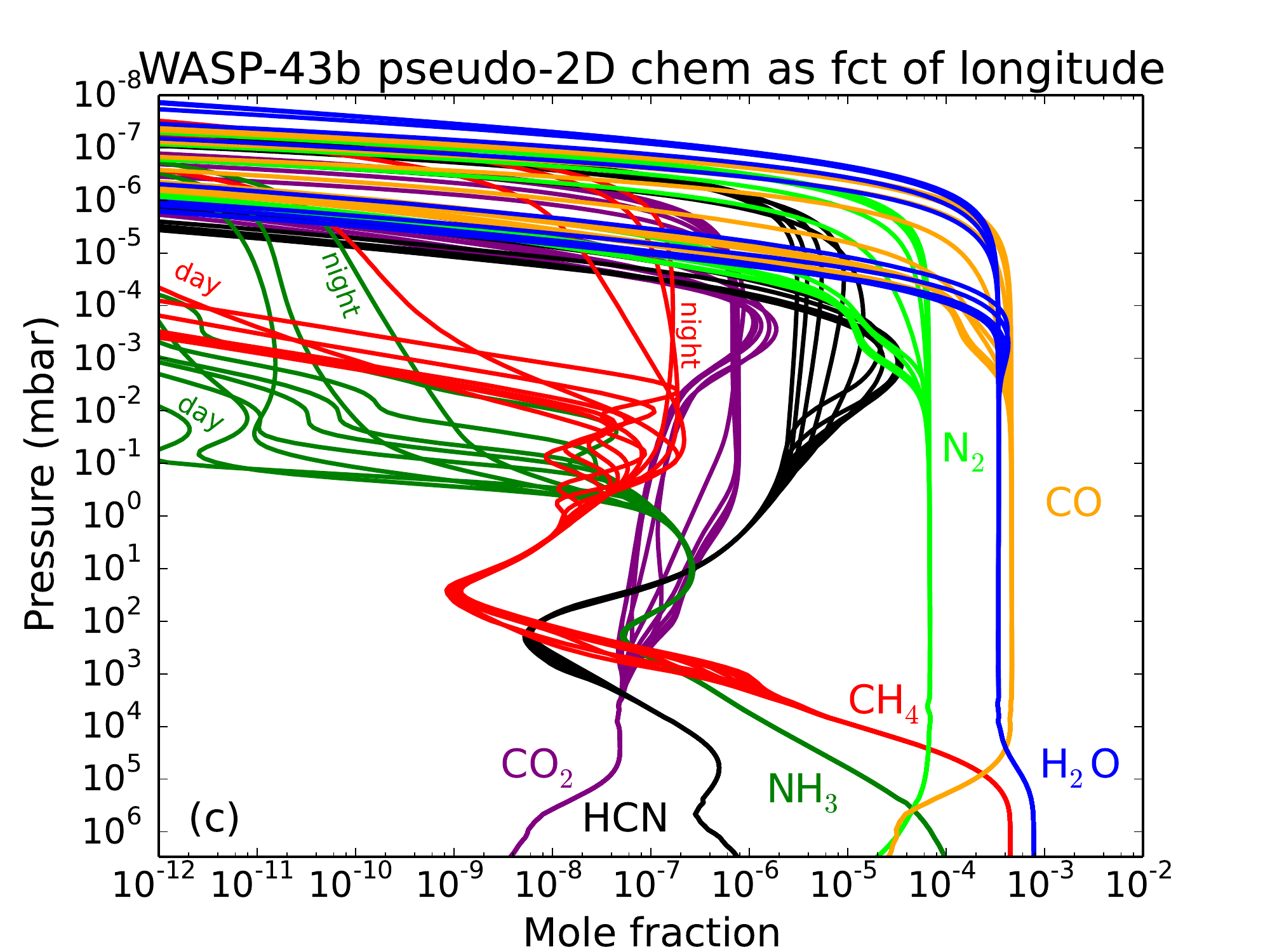}\\

\caption{Mixing ratio profiles for important atmospheric constituents on WASP-43b (as labeled) at 10 different longitudes across the planet (every 36 degrees) from (a) a model that assumes thermochemical equilibrium, (b) our 1D thermo/photochemical model that tracks chemical kinetics and vertical transport, or (c) our pseudo-2D model that tracks in addition horizontal transport.}
\end{figure}

We study the chemical composition of WASP-43b at different longitudes with our 1D and pseudo-2D models. The results are presented in Fig.~\ref{fig:abun_chem}, together with the abundances at thermochemical equilibrium, corresponding to the same longitude-variable thermal structure. In the 1D model, all the longitudes have been computed independently, assuming thermochemical composition as initial composition at each longitude. The different vertical columns don't interact with each other. In the pseudo-2D model, the longitudes are not independent interacting through horizontal circulation. As we explained in Sect. \ref{sec:chemicalmodels}, the steady state composition of the substellar point is given as initial condition to the adjacent longitude. There, the evolution of chemical composition is calculated over the amount of time necessary for a parcel of gas to reach the next longitude, and so on.

In all models, as the temperature is identical at each longitude for pressures greater than $\sim$10$^3$ mbar, we find that species have also the same abundances, corresponding to the thermochemical equilibrium values. The composition varies with longitude above this region, more or less depending on species.
For pressures lower than $\sim$10$^3$ mbar, many of the atmospheric constituents would vary significantly with longitude if the atmosphere remained in thermochemical equilibrium throughout. Particularly noteworthy is the fact that CH$_4$ would be the dominant carbon-bearing constituent at high altitudes on the colder nightside in thermochemical equilibrium, while CH$_4$ would virtually disappear from the dayside and CO would become the dominant carbon-bearing constituent at all altitudes.
The 1D kinetic model predicts that vertical quenching will reduce this variation, but there are still several orders of magnitude differences between the abundances of the dayside and that of the nightside.
In contrast, the pseudo-2D model predicts much less variation with longitude, particularly in the 0.1--1000 mbar region that is probed at infrared wavelengths. The CO that forms on the hot dayside cannot be chemically converted to CH$_4$ quickly enough on the nightside, before the atmospheric parcels are carried by the zonal winds back to the dayside. These results confirm the findings by~\citet{cooper2006}, \citet{agundez2014aa}, \citet{Mendonca2018chemistry}, \citet{Drummond2018} and \citet{Drummond2018a} that both vertical and horizontal chemical quenching are important in hot Jupiter atmospheres.

Another species whose abundance predicted by our kinetic model is very different from what is expected by thermochemical equilibrium is HCN. At thermochemical equilibrium, this species has the same abundance profile at each longitude, and its abundance decreases with increasing altitude. The 1D kinetic model predicts that this species will be quenched at around 10$^2$ mbar, leading to a higher abundance than what is predicted by thermochemical equilibrium on the nightside, and even higher abundance on the dayside thanks to a photochemical production. Note that the quenching pressure we determine with our model is of course highly dependent on the $K_{zz}$ profile we assume. At 10$^2$ mbar, $K_{ZZ}$ is about 4.5$\times$10$^7$ cm$^2$ s$^{-1}$. As we said in Sect. \ref{sec:chemicalmodels}, this parameter is rather uncertain and could vary by several orders of magnitude (typically 10$^6$--10$^{12}$ cm$^2$ s$^{-1}$ among \citealt{parmentier2013,agundez2014aa}). Consequently, with these extreme values, the pressure level quenching of HCN could vary between 10 and 10$^5$ mbar.
In contrast to what has been found with the 1D kinetic model, our pseudo-2D model indicates that the abundance of HCN on the nightside will remain very high and close to that of the dayside thanks to the horizontal circulation, in agreement with \cite{agundez2014aa}. A such high abundance might be detectable thanks to high-resolution spectroscopic observations in the near-infrared coupled to a robust detrending method \citep{hawker2018,cabot2019}. On \jwst/MIRI observations, HCN could eventually appear in the 7--8 $\mu$m band, albeit spectra will probably be dominated by water absorption in this region given the important abundance of H$_2$O in the atmosphere of WASP-43b \citep{rocchetto2016}.

Similarly to \cite{agundez2014aa}, we find that in addition to vertical quenching due to eddy diffusion, the horizontal circulation leads to horizontal quenching of chemical species. Globally, the atmosphere of WASP-43b has a chemical composition homogenized with longitude to that of the dayside. This is particularly true for pressures larger than 1 mbar, while variations of abundances between the day and nightside still remain at lower pressures. 

In summary, the pseudo-2D model suggests that CH$_4$ would be a relatively minor constituent on WASP-43b at all longitudes, that photochemically produced HCN will be more abundant than CH$_4$ in the infrared photosphere of WASP-43b at all longitudes, and that the key spectrally active species H$_2$O and CO will not vary much with longitude on WASP-43b. Benzene (C$_6$H$_6$) is a proxy for photochemical hazes in the pseudo-2D model, and the strong increase in the benzene abundance at nighttime longitudes suggests that refractory hydrocarbon hazes could potentially be produced at night from radicals produced during the daylight hours \citep[e.g.][]{miller-ricci2012,morley2013,morley2015}. Note that a recent experimental study demonstrates that refractory organic aerosols can be formed in hot exoplanet atmospheres with a C/O ratio higher than solar \citep{fleury2019}.

Based on these chemical models and because the variation of CH$_4$ with longitude could be observed with MIRI, we ran GCMs assuming chemical equilibrium and assuming a fixed [CH$_4$]/[CO] ratio of 0.001, which is representative of the 2D chemical model in the 0.1-1000 mbar region.

\subsection{Cloud coverage}\label{sec:results_clouds}

 \begin{figure}[!h]
 \label{fig:cloud_tau_reff}
 \includegraphics[width=\columnwidth]{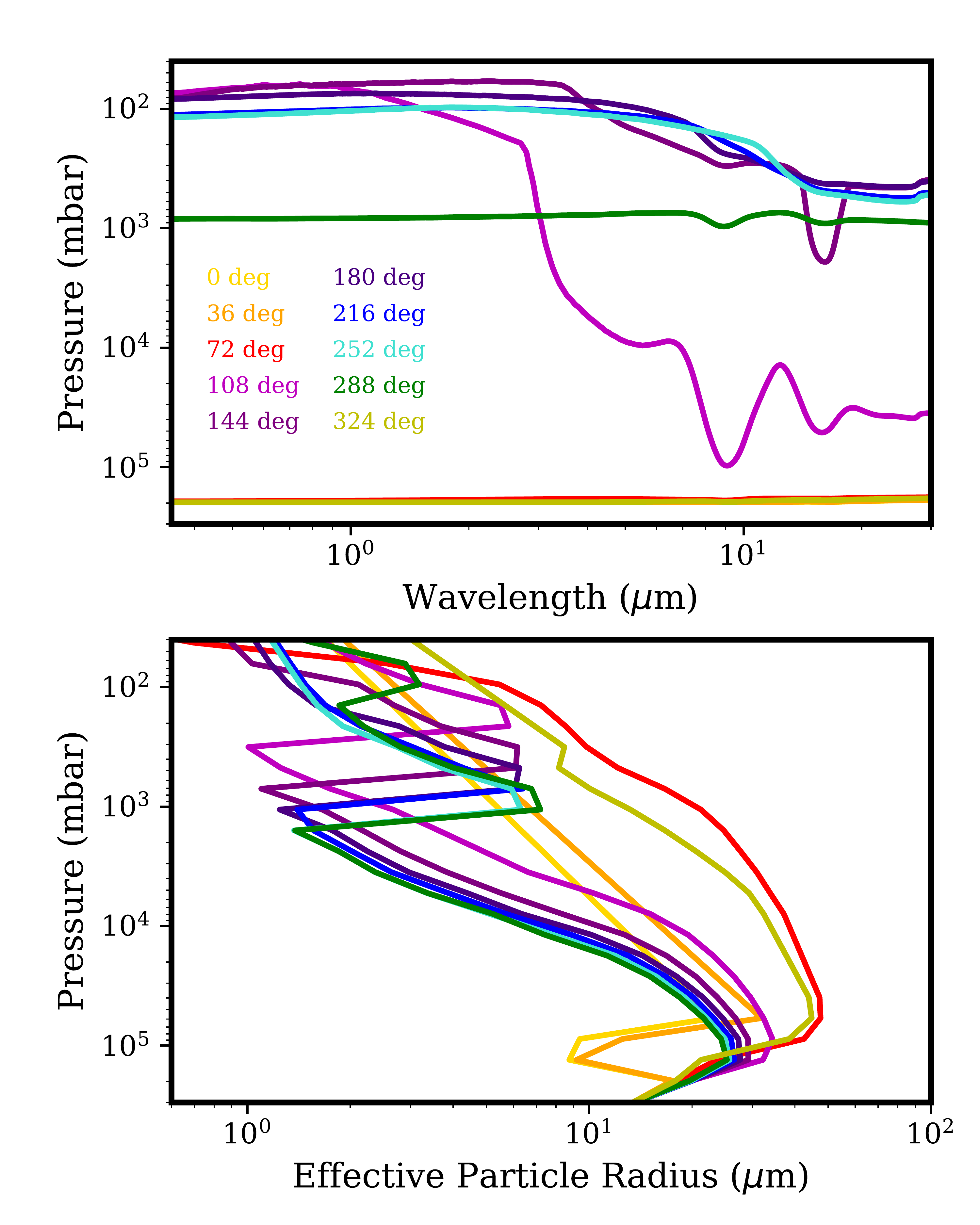}
 \caption{Pressure levels where the total cloud column optical depth equals 1 (top) and the effective particle radius profiles (bottom) predicted by \texttt{CARMA} for the labeled longitudes. The effective particle radius is calculated by averaging the size distributions of the individual cloud species; the actual mean particle radii of each species range from 0.01 to 100 $\mu$m and are functions of altitude (pressure).}
 \end{figure}
 
\begin{figure*}
 \label{fig:cloud_comp}
 \includegraphics[width=\textwidth]{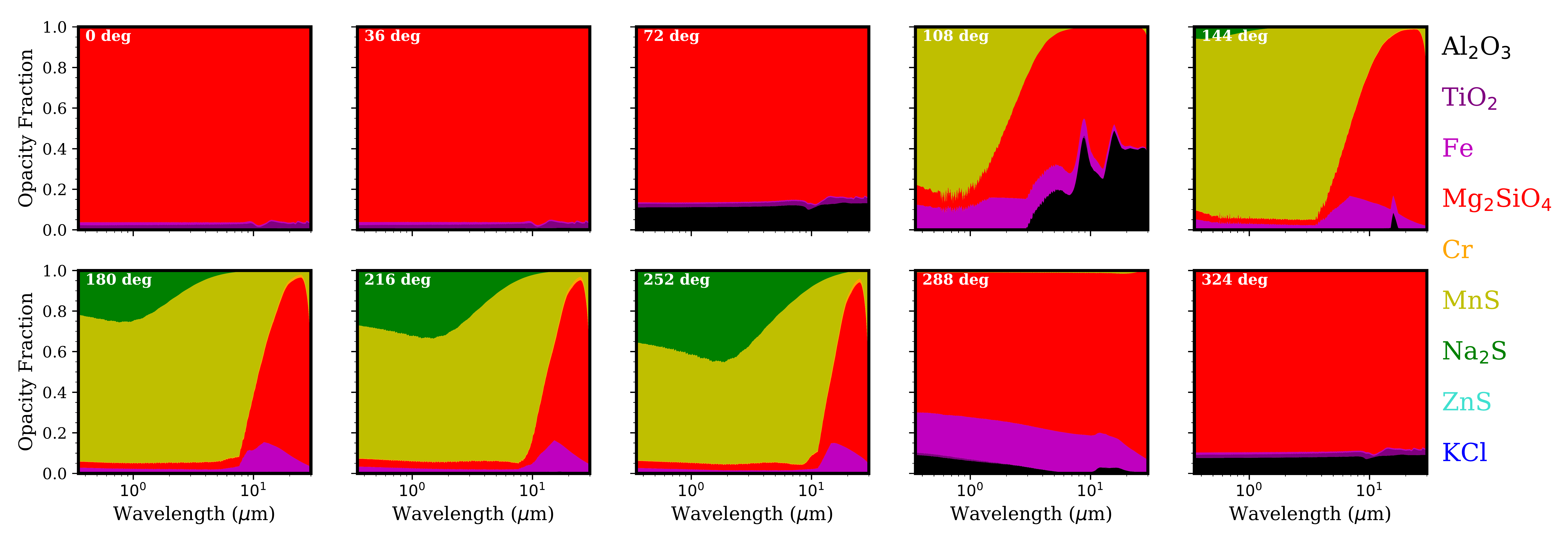}
 \caption{Contributions to the optical depth from each simulated cloud species at the pressure level where the total cloud column optical depth equals 1, as computed by \texttt{CARMA} for the labeled longitudes. Each cloud species is constituted in homogeneous particles except for forsterite and Fe, which have TiO$_2$ cores.}
 \end{figure*}

 We use the CARMA model to determine the physical and chemical properties of the clouds along different vertical columns at the equator of the planet. Assuming Mie-scattering particles, we calculate the cloud optical depth profile for each longitude from 0.35 to 30 $\mu$m. Fig.~\ref{fig:cloud_tau_reff} (top) shows the pressure levels at which the total cloud column optical depth (taking into account all cloud species) equals 1, and reveals large differences between the day and night sides. Specifically, the day side temperature profile is such that most of the forsterite (Mg$_2$SiO$_4$) is cold trapped below 100 bars, and although the forsterite condensation curve crosses the temperature profile again at lower pressures, the abundance of Mg there is sufficiently low so as to prevent optically thick clouds from forming (Fig.~\ref{fig:cloud_comp}). 
 
On the night side, sufficiently low temperatures allow for the condensation of optically thick MnS and Na$_2$S clouds, such that the optical depth = 1 pressure level is above 0.1 bar blueward of 7 $\mu$m. As the typical particle sizes of these clouds are 1 to a few $\mu$m (Fig.~\ref{fig:cloud_tau_reff}, bottom), they become optically thin at longer wavelengths, allowing for forsterite clouds to become visible, as shown by the 10 $\mu$m silicate feature. Note that this forsterite cloud is not the cold-trapped cloud at 100 bars. Instead, because the forsterite condensation curve crosses the temperature profile a second time at higher pressures here than on the day side, there is sufficient Mg to produce an optically thick ``upper'' cloud even after accounting for cold trapping. The shape of the effective particle radius profiles shown in Fig.~\ref{fig:cloud_tau_reff} (bottom) also reveals this transition in cloud composition with longitude, as particle size tends to increase towards the cloud base due to available condensate vapor supply and size sorting by lofting and sedimentation. For example, while day side profiles are largely smooth, corresponding to the dominance of the forsterite cloud, the night side profiles feature a MnS cloud deck above 1 bar sitting atop the forsterite cloud below (Fig.~\ref{fig:cloud_comp}). 

Our results suggest that whether forsterite or enstatite is considered the primary silicate condensate could strongly impact the dayside cloud opacity. As enstatite condenses at lower temperatures (Fig.~\ref{fig:temp2D}), it would not form a deep cloud at pressures~$>$~100 bars on the day side, like forsterite. This lack of cold trapping may result in an optically thick cloud at lower pressures. This is in contrast to the nightside, where forsterite only dominates the cloud opacity at long wavelengths. We therefore expect that, since the cloud base of forsterite is only $\sim$50\% higher in pressure than enstatite, forsterite will have similar effect on the night side spectra as enstatite. Whether the forsterite or enstatite clouds are cold trapped in the deep atmospheric layers depends on both microphysical behavior of the cloud (studied here), the strength of the vertical mixing and the temperature in the deep atmosphere~\citep{powell2018}. \citet{thorngren2019} recently predicted a connection between planet equilibrium temperature and their intrinsic flux, suggesting that cold traps on certain hot Jupiters may not exist due to high temperatures in the deep atmosphere. By determining the cloud chemical composition in the nightside of WASP-43b through our \jwst/MIRI phase curve observation will provide insights into the presence of a deep cold trap and thus test the predictions from \citet{thorngren2019}. %

Our work decouples cloud microphysics from the radiation field and dynamics of the rest of the atmosphere, and thus we cannot treat cloud radiative feedback or cloud advection.  Fully coupled 3D models that include cloud microphysics in the form of grain chemistry have been applied to other individual exoplanets in the past, including HD 189733b and HD 209458b, which have similar temperatures to WASP-43b \citep{helling2016,Lee2016,Lines2018a,Lines2018}. These works show that the mean particle radii vary between 1-100 $\mu$m between 0.1 and 100 bars, and that the composition of mixed cloud particles is dominated by enstatite, forsterite, iron, SiO, and SiO$_2$, with forsterite being more abundant than enstatite at most longitudes. This is similar to our results, though we do not consider SiO and SiO$_2$ in our model, while they do not consider sulfide clouds in theirs. Advection tends to smooth out cloud composition differences, which we do not capture in our work. One other major difference between the grain chemistry models and our model is the high abundance of small particles at low pressures in grain chemistry models stemming from high nucleation rates at low pressures. In contrast, nucleation rates are the highest at the cloud base in our model \citep{gao2018} owing to the high atmospheric density there, and so we lack a low pressure, small particle population.

To summarize, our results show that, if silicates primarily form forsterite clouds, then the day side of WASP-43b should be cloudless down to 100 bars, while the night side cloud opacity should be dominated by MnS and Na$_2$S clouds shortward of 7 $\mu$m, and forsterite clouds at longer wavelengths. Cloud particle sizes on the night side at the pressure levels where clouds become opaque are on the order of 1 to a few $\mu$m. On the other hand, if silicates primarily form enstatite clouds, then the dayside should be cloudier at pressures $<$100 bars, while the nightside cloud opacity would remain dominated by the sulfide clouds. 

\subsection{3-D thermal structure}\label{sec:results_3D}

\begin{figure}
 \label{fig:3Dtemp_comp}
 \includegraphics[width=\columnwidth]{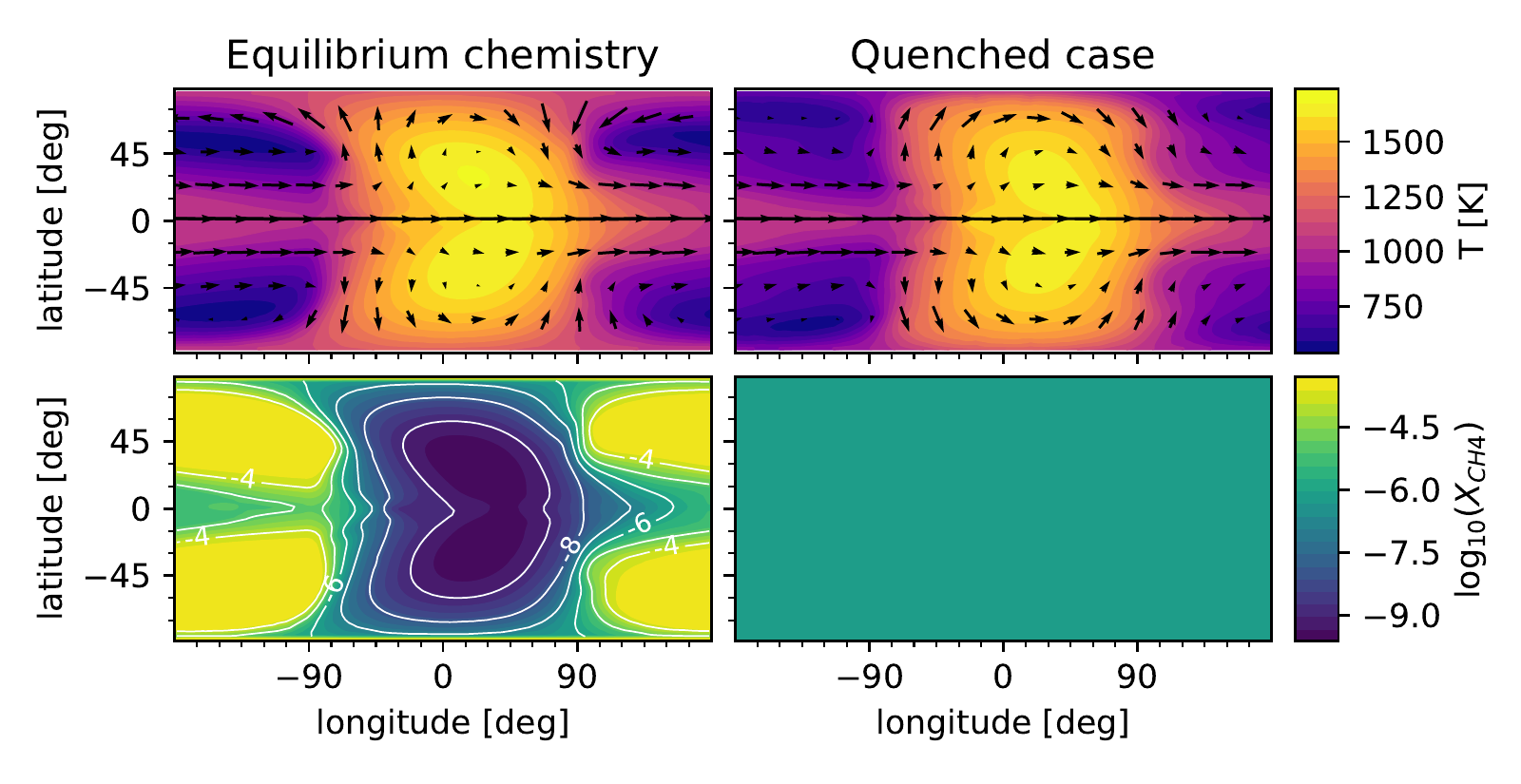}
 \caption{Temperature (top) and methane abundance (bottom) at the 30 mbar level from our 3D simulations. The substellar point is at $0^\circ$ longitude. The simulation assuming thermochemical equilibrium is shown to the left and the simulation assuming quenched CH$_4$, CO and H$_2$O abundances is shown to the right.}
 \end{figure}

We use our 3D model to calculate the thermal structure of WASP-43b assuming different chemical composition (thermochemical equilibrium and disequilibrium) and cloudy conditions (clear, MnS, and MgSiO$_3$). The temperature structure and CH$_4$ abundances for the cloudless chemical equilibrium and disequilibrium simulations are shown in Fig. \ref{fig:3Dtemp_comp}. From these models, we calculate the corresponding emission spectra at dayside and nightside.

\begin{figure}[!h]
\label{fig:3dSpec}
\includegraphics[width=\columnwidth]{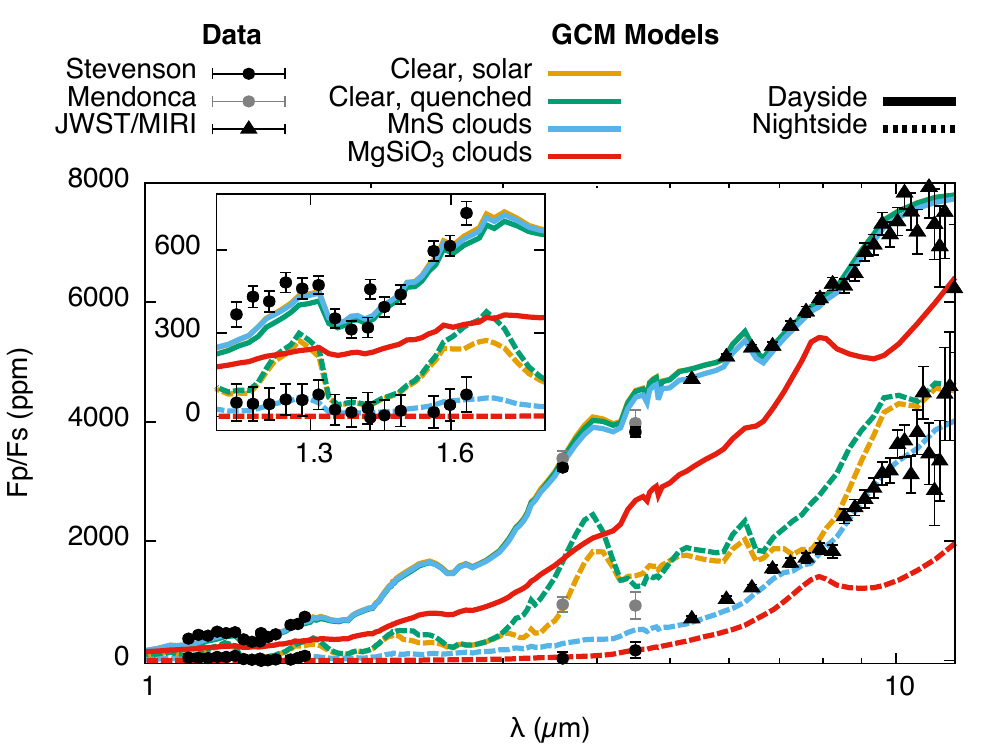}
\caption{Dayside (plain) and nightside (dashed) spectrum of WASP-43b predicted by the \texttt{SPARC/MITgcm} for cloudless, cloudy (with 1 $\mu$m particles) or non-equilibrium assumptions. Current \hst and \textit{Spitzer} \citep{stevenson2017,Mendonca2018} observations are shown as dots, planned \jwst/MIRI observations are shown as triangles.}
\end{figure}

 \begin{figure}[!h]
 \label{fig:3Spec_size}
 \includegraphics[width=\columnwidth]{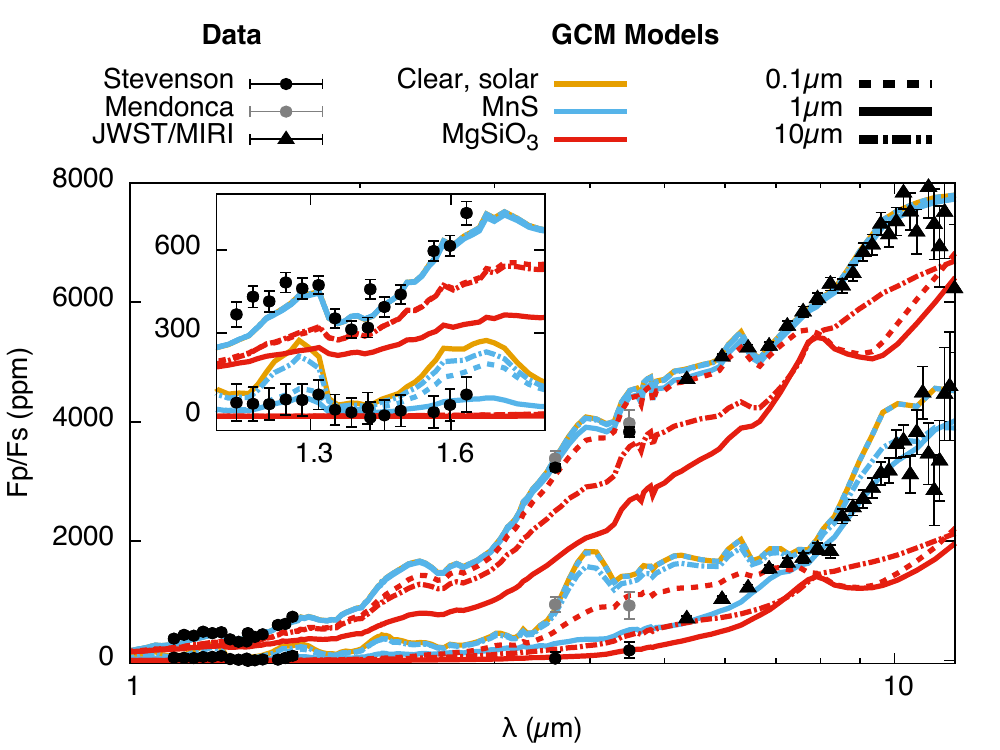}
 \caption{Dayside (top curve) and nightside (bottom curves) spectrum of WASP-43b predicted by the \texttt{SPARC/MITgcm} for our cloudless models and models with passive MnS (blue) and $\rm MgSiO_3$ (red) clouds of different particle sizes. Current \hst and \textit{Spitzer} observations are shown as dots, planned \jwst/MIRI observations are shown as triangles.}
 \end{figure}

 \begin{figure}[!h]
 \label{fig:RadFeedback}
 \includegraphics[width=\columnwidth]{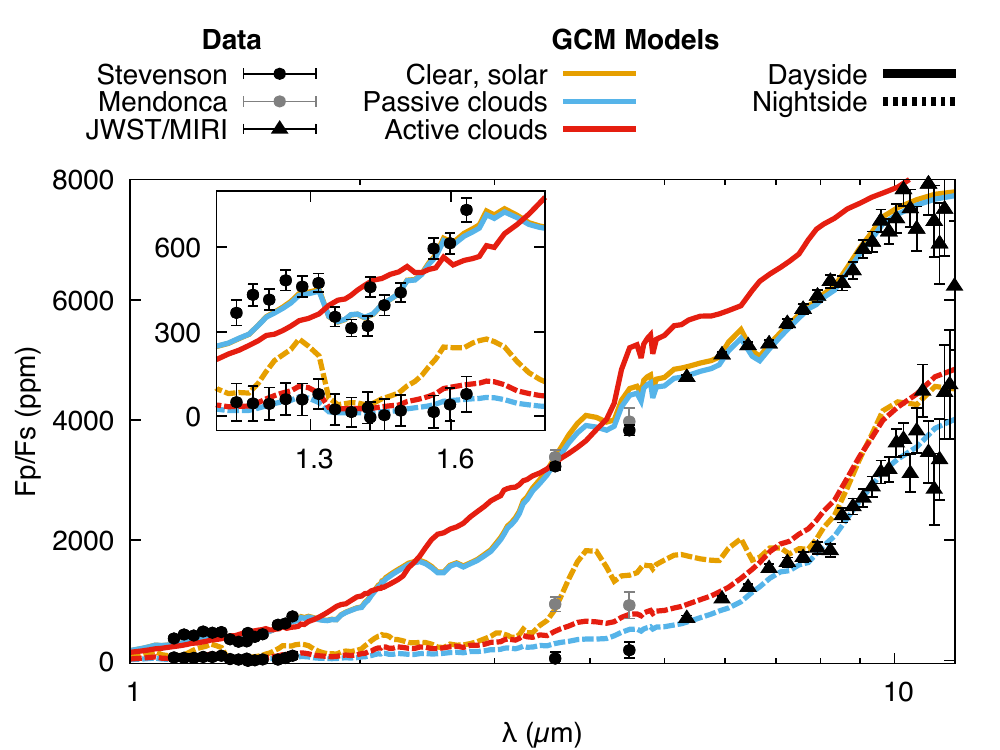}
 \caption{Dayside (top curve) and nightside (bottom curves) spectrum of WASP-43b predicted by the \texttt{SPARC/MITgcm} for our cloudless models, models with passive MnS, $1\mu m$ clouds (blue) and models with radiatively active MnS, $1\mu m$ clouds (red). Current \hst and \textit{Spitzer} observations are shown as dots, planned \jwst/MIRI observations are shown as triangles.}
 \end{figure}
 
The thermal structure of our cloudless, chemical equilibrium \texttt{SPARC/MITgcm} simulations are very similar to the one presented in~\citet{kataria2015} where the reader can find a thorough description of the atmospheric flows. While~\citet{kataria2015} focused on the effect of TiO, metallicity and drag, we hereafter discuss the role of disequilibrium chemistry and clouds in shaping the nightside spectrum of the planet. 

In the case of quenched carbon chemistry ([CH$_4$]/[CO] = 0.001), the dayside is slightly cooler and the nightside is slightly warmer at a given pressure level than for our chemical equilibrium case. However, the differences in the spectra seen in Fig.~\ref{fig:3dSpec} are mainly due to change in the opacities rather than changes in the thermal structure. On the dayside, where the [CH$_4$]/[CO] ratio is small at chemical equilibrium, our quenched and  chemical equilibrium simulations are indistinguishable. On the nightside, the quenching removes the CH$_4$ absorption bands between 3 and 4 $\mu$m and the ones between 7 and 9 $\mu$m are weakened in the quenched scenario, leading to a signature detectable by \jwst/MIRI. Note that our GCM simulations approximate the [CH$_4$]/[CO] ratio to be constant throughout the atmosphere (horizontally and vertically) for computational reasons, while the pseudo-2D simulations in Section \ref{sec:results_chemistry} as well as 3D simulations of WASP-43b with a simplified chemistry scheme \citep{Mendonca2018chemistry} find that the methane abundance is homogenized horizontally but decreases with increasing altitude. However, \citet{steinrueck2018} found that the effect of disequilibrium chemistry on the thermal structure and phase curve is qualitatively similar for different constant [CH$_4$]/[CO] ratios as long as CO is the dominant carbon-bearing species. Therefore, it is likely that the effect of a horizontally homogenized [CH$_4$]/[CO] ratio that decreases with altitude is also qualitatively similar. Our quenched simulation thus still provides a valuable estimate of the effects of disequilibrium chemistry. 

The cloudless simulations were also post-processed with cloud opacities. The post-processing allows for a quick estimate of the strength of potential signature of cloud properties in the emission spectrum without the need to run additional, time-consuming, global circulation models. In Sect.~\ref{sec:results_clouds}, we found that the nightside of WASP-43b could be dominated by MnS, Na$_2$S, MgSiO$_3$, and/or Mg$_2$SiO$_4$. Following~\citet{Parmentier2016}, we explore two possible cloud compositions: MnS and MgSiO$_3$. Forsterite and enstatite having opacities and condensation curves very similar, we chose to include just one of these silicate species.  As we found with our microphysical cloud model (Sect. \ref{sec:results_clouds}), the atmosphere of WASP-43b is cool enough for MgSiO$_3$ clouds to cover the whole planet affecting both the dayside and the nightside of the planet. Conversely, MnS clouds can only form on the cooler nightside and thus only affect the nightside's spectrum. Both MnS and MgSiO$_3$ clouds are able to sufficiently dim the nightside emission spectrum blueward of 5 $\mu$m in order to match the \hst and the \textit{Spitzer} observations. As shown in Fig.~\ref{fig:3dSpec}, in all our models the nightside flux remains observable with \jwst/MIRI, even when the thermal emission is extremely small shortward of 5 $\mu$m. MnS and MgSiO$_3$ cloud composition could be distinguished spectrally by our \jwst/MIRI phase curve observation through the observation of the 10 $\mu$m absorption band seen in the red models of Fig.~\ref{fig:3dSpec} \citep[see also][]{Wakeford2015}.

The effect of the cloud particle size is explored in Fig.~\ref{fig:3Spec_size}. Assuming that the formation of MnS clouds is limited by the available amount of manganese in a solar-composition atmosphere, the MnS clouds could be either transparent or optically thick in the \jwst/MIRI bandpass depending on the size of their particles. Conversely, $\rm MgSiO_3$, if present, should always be optically thick in the MIRI bandpass. 

The radiative feedback effect of the clouds in hot Jupiter is a subject of intense research. The amplitude and spatial distribution of the cloud heating is extremely dependent on the cloud model used~\citep[e.g.][]{Lee2016,Roman2017,Lines2018,Lines2019,Roman2019}. In Figure~\ref{fig:RadFeedback} we show the resulting spectrum from a global circulation model incorporating the radiative feedback of MnS clouds. The clouds are opaque up to mbar pressures on the planet nightside and produce a strong greenhouse effect, leading to a warmer nightside and thus a higher nightside flux. Horizontal heat transport from nightside to dayside changes the dayside thermal structure by a thermal inversion, leading to a dayside spectrum dominated by emission features. Qualitatively, any nightside clouds should increase the nightside opacity and warm the atmosphere. The lower the pressure of the cloud photosphere the higher the greenhouse effect of the clouds should. As a consequence, the dayside photosphere, should warm up through heat transport from the nightside to the dayside. The lower the photospheric pressure on the nightside, the larger the warming effect of the clouds on the dayside photosphere.  The exemple shown here assumes the highest possible cloud and is therefore likely to overestimate the effect of the nightside cloud on the dayside thermal structure. A deeper cloud will likely have a smaller impact on the dayside spectrum. Overall, because the amplitude and spatial distribution of the cloud heating is extremely dependent on the cloud model used~\citep[e.g.][]{Lee2016,Roman2017,Lines2018,Lines2019}, we decided to focus the reminder of the paper on the post-processed case by comparing the spectral effect of the clouds for a given thermal structure.



The (cloud-free) quenched simulations are not a good match to existing \hst and \textit{Spitzer} nightside observations. Most likely, this is because the effect of nightside clouds dominates over the effect of disequilibrium chemistry at the wavelengths of existing observations. This situation is qualitatively similar to what \citet{steinrueck2018} find for HD~189733b. However, quenched chemistry might still be important on the nightside of WASP-43b. 
We note that during the referee process of this paper, \cite{morello2019} published a new reduction of the \textit{Spitzer} observations of WASP-43b. The resulting nightside fluxes lay between those of \cite{stevenson2017} and \cite{Mendonca2018}.

Also, we note that a new 3D circulation model, elicits for WASP-43b a dynamical regime different from \texttt{SPARC/MITgcm}. \cite{Carone2019} propose that the presence of very deep wind jets down to 700 bar leads to an interruption of superrotation and thus also an interruption of day-to-night side heat transfer. In contrast to that, our \texttt{SPARC/MITgcm} model does not display very deep wind jets. It has uninterrupted superrotation and thus an efficient day-to-night side heat transport. Thus, the \cite{Carone2019}'s model yields colder night sides by several 100 K compared to our model (see also Fig.7 in \citealt{Carone2019} for a direct comparison).

\section{\jwst simulations}\label{sec:JWST_simu}

We run the \jwst simulations following the procedure and assumptions described in Section \ref{sec: JWST Observation Model} using the different GCM spectra as inputs. 
Our models predict that there should be clouds on the night-side of WASP-43 b, but other models are consistent with cloud-free night-side atmospheres with deep winds or drag \citep{Carone2019, Komacek2016}. Thus, we perform simulations for cloudy cases and for the cloud-free (``clear'') case to investigate whether they can be distinguished: the cloud-free case should be rejected by the data if the atmosphere is indeed cloudy. Beyond the case study of WASP-43 b, these simulations can inform \jwst observation programs of other hot Jupiters, which may have cloudy or clear atmospheres.

The timing and exposure parameters are optimized within \texttt{PandExo}. For these simulations (WASP-43, K\textsubscript{mag} = 9.27), the computed parameters are 0.159 second per frame, one frame per group, 83 groups per integration, for a total of 13.36 seconds per integration. This yields 293 integrations during each one eighteenth of the phase curve and 627 in-eclipse integrations including both eclipses. The observing efficiency is 98\%. We consider only the 5 -- 12 $\mu$m spectral range. For the in-eclipse observations, the mean electron rate per resolution element at the native MIRI LRS resolution is 5168 e$^-$/s, the median is 2725 e$^-$/s, and it varies from 25229 to 395 e$^-$/s from 5.4 to 12 $\mu$m. This corresponds to a signal to noise ratio of 13333 at 5.4 $\mu$m and 536 at 12 $\mu$m per resolution element. No warnings were issued during the simulations. As expected, these exposure parameters differ slightly from the final ones that are obtained with the \jwst Exposure Time Calculator (ETC) and the Astronomer's Proposal Tool (APT), but this does not affect our results. 

\begin{figure}[!h]
   \centering
   \includegraphics[width=\columnwidth]{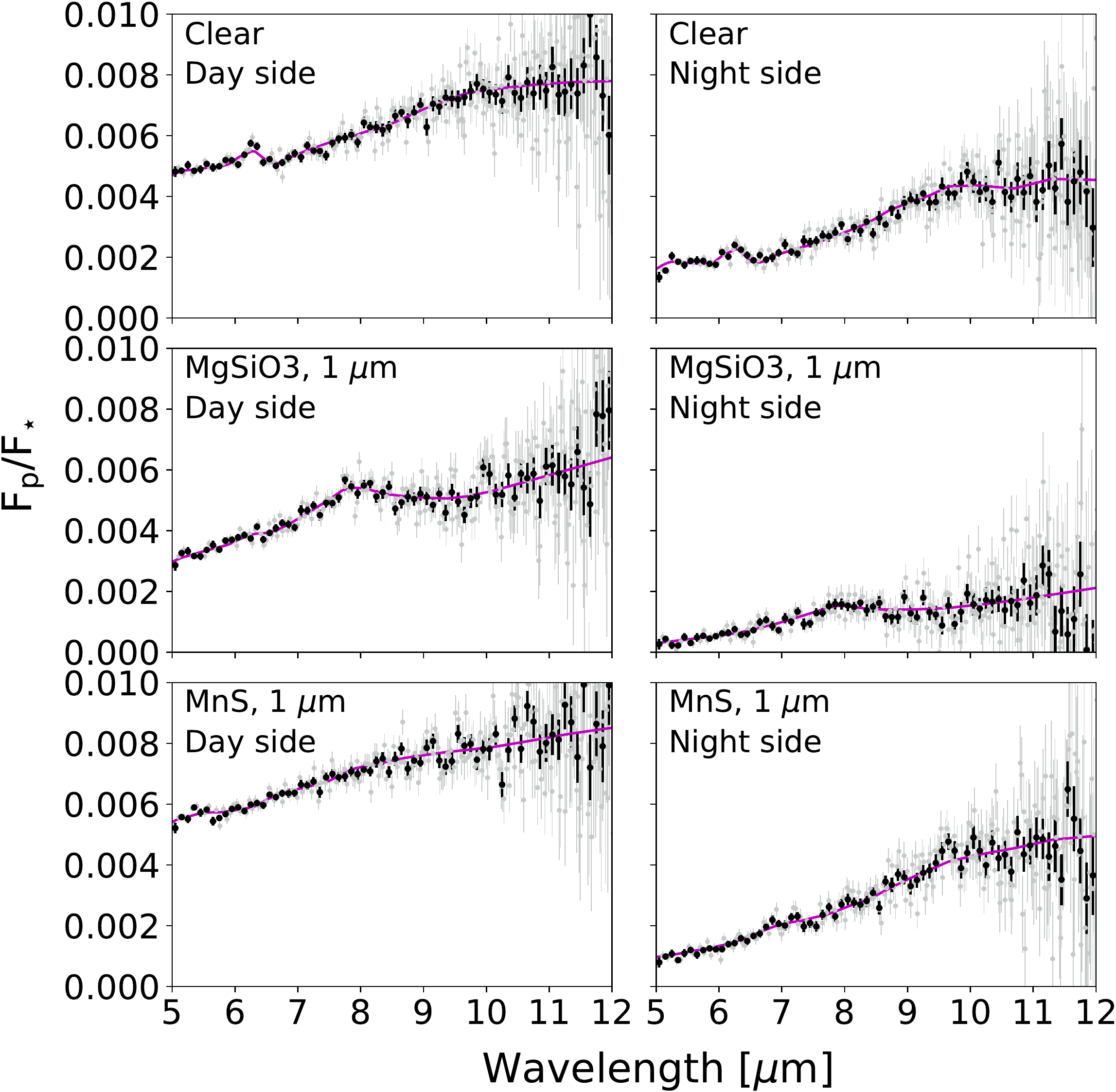}
   \caption{\texttt{PandExo} simulations of the emission spectrum of WASP-43b observed with MIRI LRS, for a model with a clear atmosphere and quenching (top), a model with MgSiO$_3$ clouds and 1 $\mu$m particle size (middle), and a model with MnS clouds and 1 $\mu$m particle size (bottom), for the day side (left) and the night side (right). The simulated data at the native MIRI LRS resolution are shown in gray with their 1$\sigma$ uncertainties. The same data and uncertainties averaged into equal wavelength bins of 0.1 $\mu$m width are shown in black. The theoretical input model spectra from \texttt{SPARC/MITgcm} are shown in magenta.}
   \label{fig: pandexo simulations}
\end{figure}

Examples of simulations are shown in Fig. \ref{fig: pandexo simulations}. At the MIRI LRS native resolution, the wavelength interval between points varies from 0.08 to 0.019 $\mu$m from 5 to 12 $\mu$m. We resample the spectra to equal wavelength bins of 0.1 $\mu$m width. The median uncertainty per spectral bin is 210 ppm, with a notable difference below and above 10 $\mu$m (with a median uncertainty of 170 ppm and 640 ppm, respectively). Adding a systematic noise floor of 50 ppm \citep{greene2016} would not significantly change these uncertainties. Taking the model with a clear atmosphere and quenching as an example, these uncertainties are smaller than the variation between the day- and the night-side by a factor of $\sim18$ and $\sim5$ below and above 10 $\mu$m, respectively. They are also smaller than the night side emission by a factor of 14 and 7 below and above 10 $\mu$m, respectively. Thus, we should be able to detect the night side and day side emission spectra and its variations in longitude. These uncertainties are also smaller than differences between models around specific spectral features, in particular the increased emission around 8 $\mu$m on the day-side for models with MgSiO$_3$ clouds should be detected. Thus, we should be able to constrain the cloud composition. A detailed retrieval analysis based on these simulations is presented in Section \ref{sec:retrieval}.

\section{Retrieval}\label{sec:retrieval}

For the atmospheric retrieval of WASP-43b, we consider that the NIRSpec GTO will constrain the {\water} mixing ratio before the ERS observations. Thus, in {\pyratbay} we consider a prior $\log(\water) = -3.52 \pm 0.3$ based on~\cite{greene2016}, while in {\taurex} we use a uniform prior (1.5--6$\times10^{-4}$ for the cloud-free retrieval, 1$\times10^{-1}$--1$\times10^{-5}$ for the cloudy ones). These priors are consistent with the water abundance determined by our chemical 2D model in the 0.1--1000 mbar region, probed by infrared observations. Since the NIRSpec GTO may not be able to constrain the atmospheric properties on the night side of the planet, we investigate whether the MIRI observations are able to determine: (1) if there are no clouds, is there a disequilibrium-chemistry composition? and (2) if there are clouds, what is their composition?

\begin{figure*}[t]
   \centering
   \includegraphics[width=0.9\textwidth]{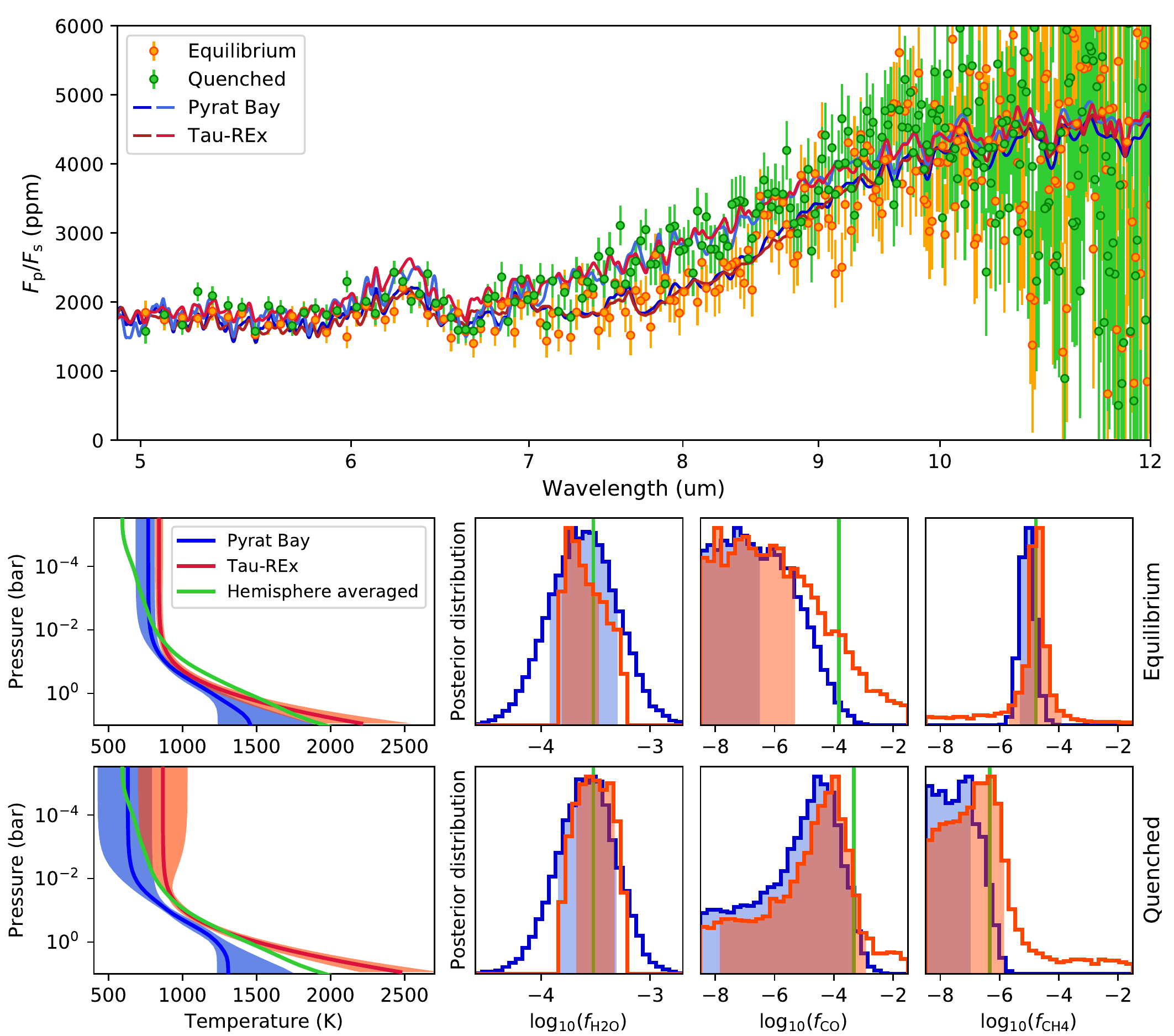}
   \caption{Cloud-free atmospheric retrieval results. The top panel shows simulated WASP-43b MIRI night-side cloud-free spectra for the equilibrium and quenched case (see legend). The dark and light solid curves denote the best-fitting model for {\pyratbay} (blue) and {\taurex} (red) of the equilibrium and quenched simulations, respectively. The middle and bottom-row panels show the model posterior distributions for the equilibrium and quenched cases, respectively. On the left panel, the blue and red solid curves denote the mean pressure--temperature profiles from the {\pyratbay} and {\taurex} posterior distributions, respectively. The shaded areas, denote the 68-percentile extent of the distributions.  The green solid curve denotes the night-side hemisphere-averaged temperature profile of the 3D input model. The panels on the right show the molecular marginal posterior distributions for {\water}, CO, and {\methane} (same color code as on the left panels). The shaded areas denote the 68\% highest-posterior-density credible interval of the respective distributions. The vertical green line denotes the hemisphere-averaged molecular mixing ratios of the input model, at 0.3~bar.}
\label{fig:retrieval-clear}
\end{figure*}

\subsection{Cloud-free Retrieval}\label{sec:cloud-free ret}

For the night-side cloud-free retrievals, both {\pyratbay} and {\taurex} reproduce well the simulated data (Fig.~\ref{fig:retrieval-clear}, top panel) and obtain similar results for the retrieved temperature profile and the molecular abundances (Fig.~\ref{fig:retrieval-clear}, bottom panels).
Note, however, that the temperatures and abundances of the individual cells in the input 3D model span wide ranges, which vary with latitude, longitude, and pressure (see, e.g., Fig. \ref{fig:3Dtemp_comp}). Since the output emission spectrum is not a linear transformation of the input temperature, one must consider the averaged values of the input model as guidelines rather than a strict measure of accuracy.

\begin{table}[!b]
\caption{Comparison of the retrieved abundances by {\taurex} and {\pyratbay}, as well as the uncertainty interval, in the cloud-free cases.} \label{tab:comp_cloudfree}
\centering
\begin{tabular}{c|c|c|c}
\hline
\hline
Case & Molecule  &      \taurex   &      \pyratbay   \\
\hline
            & log(H$_2$O) & -3.67$\pm$0.17 & -3.59$\pm$0.31\\
Equilibrium & log(CO)     & <-5.3     & < -7.0  \\
            & log(CH$_4$) & -4.82$\pm$0.89 & -5.06$\pm$0.25 \\
\hline
          & log(H$_2$O) & -3.52$\pm$0.18 & -3.62$\pm$0.27\\
Quenched  & log(CO)     & <-2.9 & < -3.3  \\
          & log(CH$_4$) & <-5.8 & < -7.2   \\
          
\end{tabular}
\end{table}

Given the properties of the system, the bulk of the emission from WASP-43b comes from the 0.1--1 bar pressure range. At these altitudes, both codes follow closely the hemisphere-averaged profile of the underlying model; fitting well the non-inverted slope of the temperature profile.

Water is the dominant absorber across the MIRI waveband. Its ubiquitous absorption at all wavelengths shapes the emission spectrum.
Both retrievals constrain well the water abundance, aided by the NIRSpec prior (Gaussian for {\pyratbay}, uniform for {\taurex}).

Methane has its strongest absorption band between 7 and 9~$\mu$m. Consequently, the larger methane abundance for the equilibrium case over the quenched case produces a markedly lower emission at these wavelengths (more methane concentration leads to stronger absorption, which leads to higher photosphere---at lower temperatures, which leads to lower emission).  Both retrievals are able to distinguish between these two cases, producing a precise methane constraint for the equilibrium case ({\pyratbay} obtains a median with 68\% HPD (highest-posterior-density) of $\log({\rm CH_4})=-5.06\pm0.25$, whereas {\taurex} obtains $\log({\rm CH_4})=-4.82\pm0.89$). The lower concentration of methane in the quenched case leads to wider posteriors and the retrieval is only able to provide an upper limit on the methane concentration.

Carbon monoxide only has a strong band at the shorter edge of the observed spectra (5~${\mu}$m), and thus, its abundance is harder to constrain. In the equilibrium case both retrievals set an upper limit on the CO abundance. In Figure \ref{fig:retrieval-clear}, the upper limits of the credible intervals are nearly two orders of magnitude below the averaged CO abundance at 0.3~bar. However, in the input model, CO decreases rapidly with altitude over the probed pressures---from $\sim$4$\times10^{-4}$ to $\sim$4$\times10^{-5}$ between 1 and 0.1~bar. Since the retrievals assume a constant-with-altitude profile, the retrieval models require a lower CO abundance to produce the same signal of the input model, which might explain the underestimated retrieved CO values. For the quenched case, both CO posterior distributions peak slightly below the averaged-input.

Table \ref{tab:comp_cloudfree} and Figure \ref{fig:retrieval-clear} compare the retrieved abundances and the associated uncertainties for the two cloud-free cases (Equilibrium and Quenched chemistry).  Both codes (based on different methods) produce qualitatively similar results for each molecule.  The posterior credible intervals are consistent when the molecule is well constrained, whereas they differ by up to two dex when finding upper limits. These results show that both methods are robust. At view of our results, we expect that in a cloud-free case we will be able to distinguish between the equilibrium and quenched scenarios using MIRI phase curve observations of WASP-43b, thanks to the methane absorption band seen between 7 and 9~$\mu$m.

\begin{figure*}[t]
   \centering
   \includegraphics[width=0.9\textwidth,trim={0 5cm 0 0}]{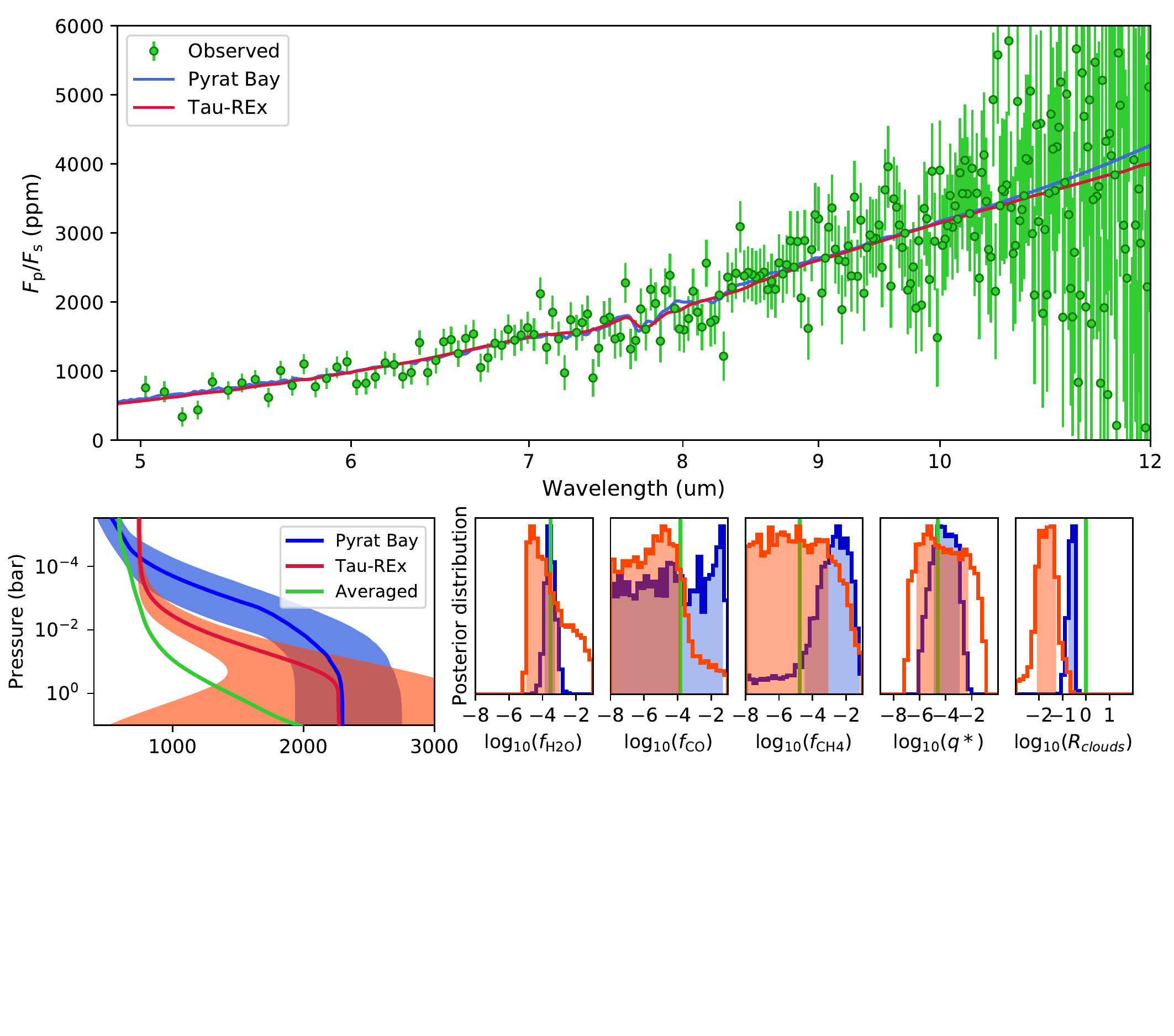}
   \caption{MnS ``free'' chemistry retrieval for \pyratbay and \taurex. In this run the $T-P$ parameters are free, and the species abundances are retrieved using free chemistry. Top panel shows the best-fit spectra with data points and uncertainties. The lower left panel shows the median $T-P$ profiles and the extent of the 1$\sigma$ regions. Lower right panel shows the retrieved posteriors.} 
   \label{fig:MnS_retrievals}
\end{figure*}


\subsection{Cloudy Retrieval}
\label{cloudy ret}

We ran {\pyratbay} and {\taurex} night-side cloudy retrievals on the 1 $\mu$m $\rm MnS$ and $\rm MgSiO_3$ {\em JWST}/MIRI simulated datasets (Figure \ref{fig: pandexo simulations}, right-middle and bottom panels). Although both codes retrieved similar best-fit spectra, temperature-profiles, and contribution functions, {\pyratbay}'s \texttt{TSC} cloud model was more successful in retrieving the input condensate particle size, cloud number density, and the location of the cloud deck for the $\rm MgSiO_3$ clouds. 

To investigate $\rm MnS$ and $\rm MgSiO_3$ clouds, {\taurex} ran ``free'' retrieval scenarios, while {\pyratbay} ran both ``free'' and ``self-consistent'' equilibrium-chemistry retrieval. We chose to include the ``self-consistent'' scenario, as the clouds in the input synthetic models were post-processed with cloud opacities using the cloudless GCM simulations, in the same way as we add clouds in retrieval (see Section \ref{sec:results_3D}). For the $\rm MgSiO_3$ clouds, both {\pyratbay} and {\taurex} also ran non-constrained and constrained temperature-profile cases within ``free'' retrieval, as the data led the parameter exploration to non-physical solutions.

\begin{figure*}[t]
   \centering
   \includegraphics[width=0.9\textwidth, trim={0 5cm 0 0}]{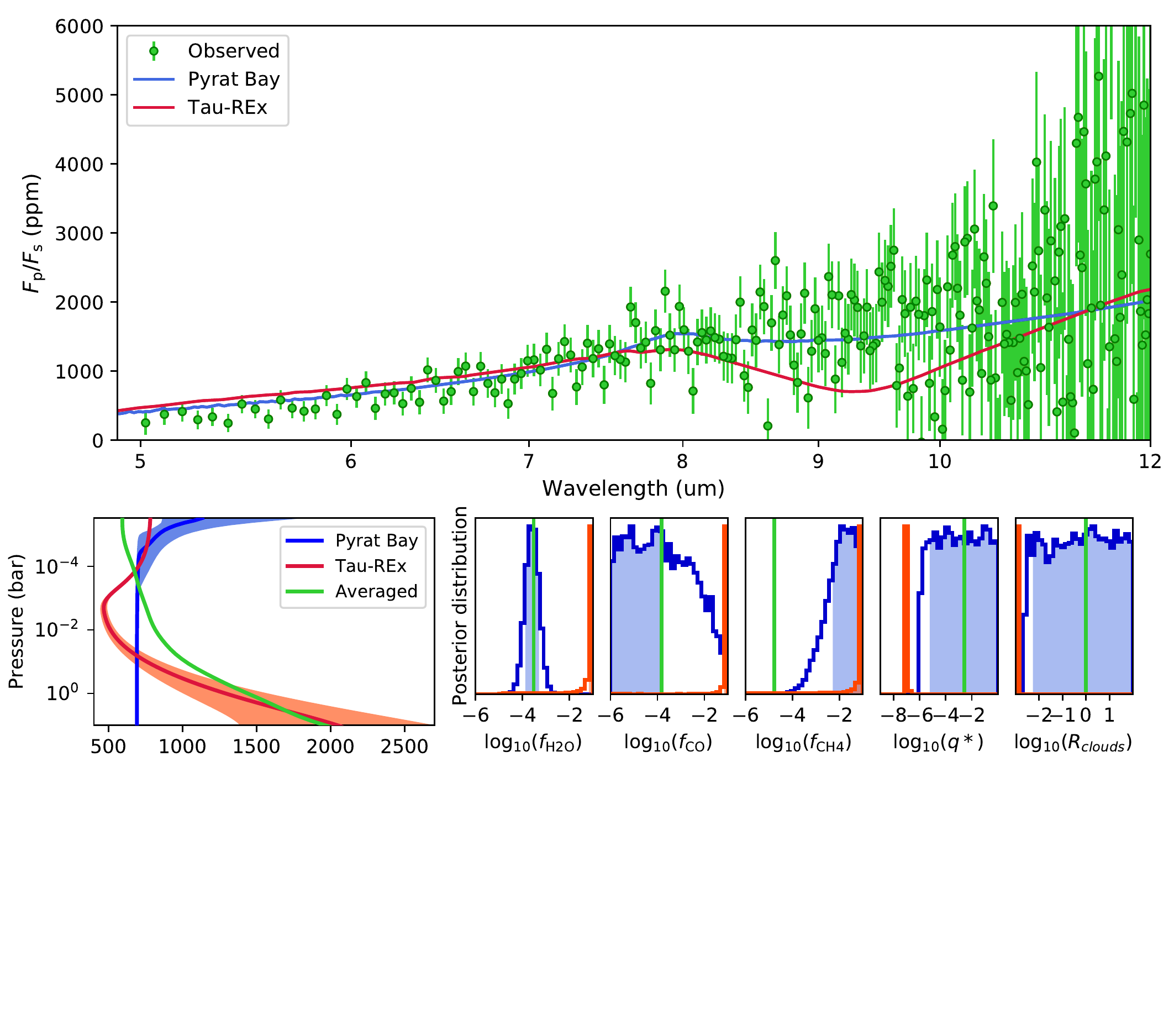}
   \includegraphics[width=0.41\textwidth]{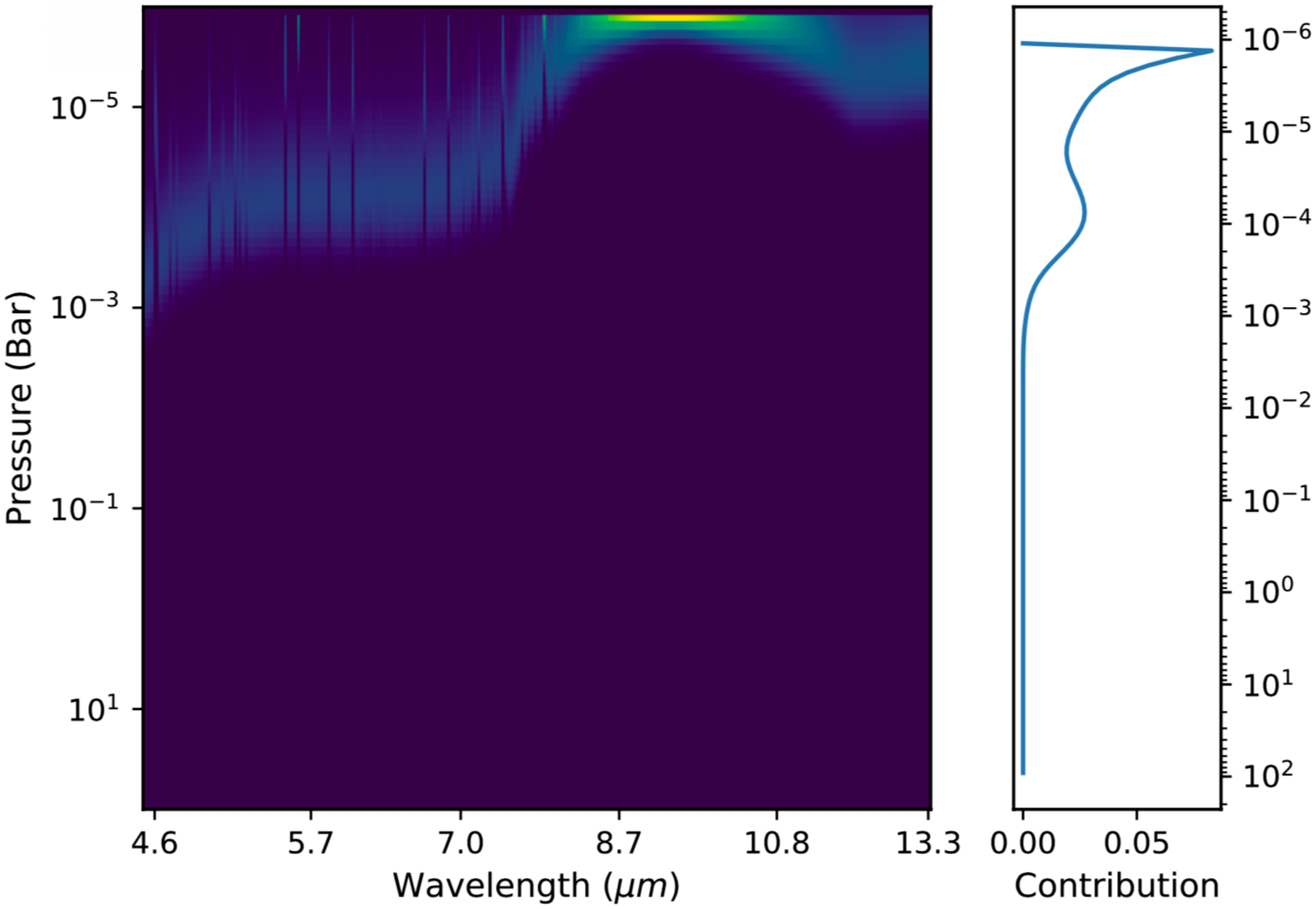}
   \includegraphics[width=0.49\textwidth]{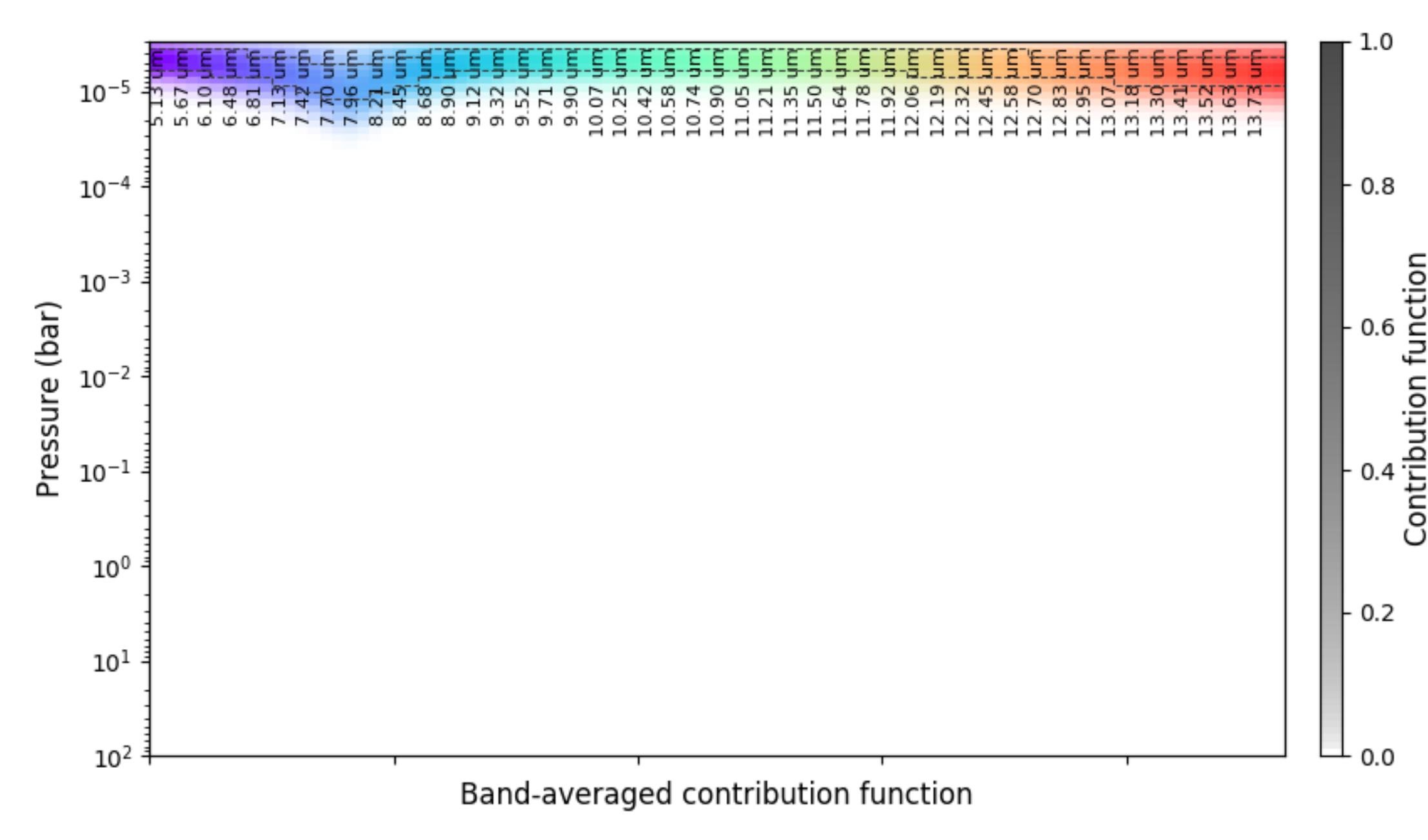}
   \caption{MgSiO$_3$ ``free'' chemistry retrieval for \pyratbay and \taurex. In this run the $T-P$ parameters are free, and the species abundances are retrieved using free chemistry. Top panel shows the best-fit spectra with data points and uncertainties. The middle left panel shows the median $T-P$ profiles and the extent of the 1$\sigma$ regions. Middle right panel shows the retrieved posteriors. The lower figures are the contributions functions from \taurex (left) and \pyratbay (right).}
   \label{fig:MgSiO3_retrievals}
\end{figure*}

For {\pyratbay} ``free'' retrieval we used the same setup as in the cloud-free retrieval case (see Sections \ref{sec:retrieval_description} and \ref{sec:cloud-free ret}) and retrieved the same temperature and pressure parameters \citep[$\kappa$, $\gamma_1$, $\gamma_2$, $\alpha$, and $\beta$; see ][]{LineEtal2013apjRetrievalI}, with an addition of five new free parameters describing the cloud characteristics (Blecic et al 2019a, in prep.): the cloud extent, $\Delta i$, the cloud profile shape, $H_c$, the condensate mole fraction, $\log_{10}(q*)$, the particle-size distribution, $\log_{10}(r_{eff})$, and the gas number fraction just below the cloud deck, $\log_{10}(X_c)$. For the ``self-consistent'' scenario we retrieved the temperature-pressure parameters together with the aforementioned cloudy parameters, excluding the chemical species parameters. The abundances of the chemical species were produced using the initial implementation of the \texttt{RATE} code \citep{CubillosBlecicDobbs-Dixon2019-RATE}, based on \citeauthor{HengTsai2016} 5-species solution. We calculated the volume mixing ratio of {\water}, {\methane}, CO, {\carbdiox}, and {\acetilen} species given a $T-P$ profile. {\taurex} uses the same setup as in Sections \ref{sec:retrieval_description} and \ref{sec:cloud-free ret}, in addition to the cloud prescription and free parameters described in \cite{bohren_1983}, Appendix A. The free parameters of the model are the condensate mole fraction, $\log_{10}(q*)$, and the peak of the particle-size distribution $\log_{10}(r_{eff})$. For this paper, we use the particle cloud distribution described in \cite{sharp_clouds}.

\begin{figure*}[t!]
   \centering
   \subfigure{\raisebox{0.6cm}{\includegraphics[height=4.6cm, trim={5cm 0 0cm 0}]{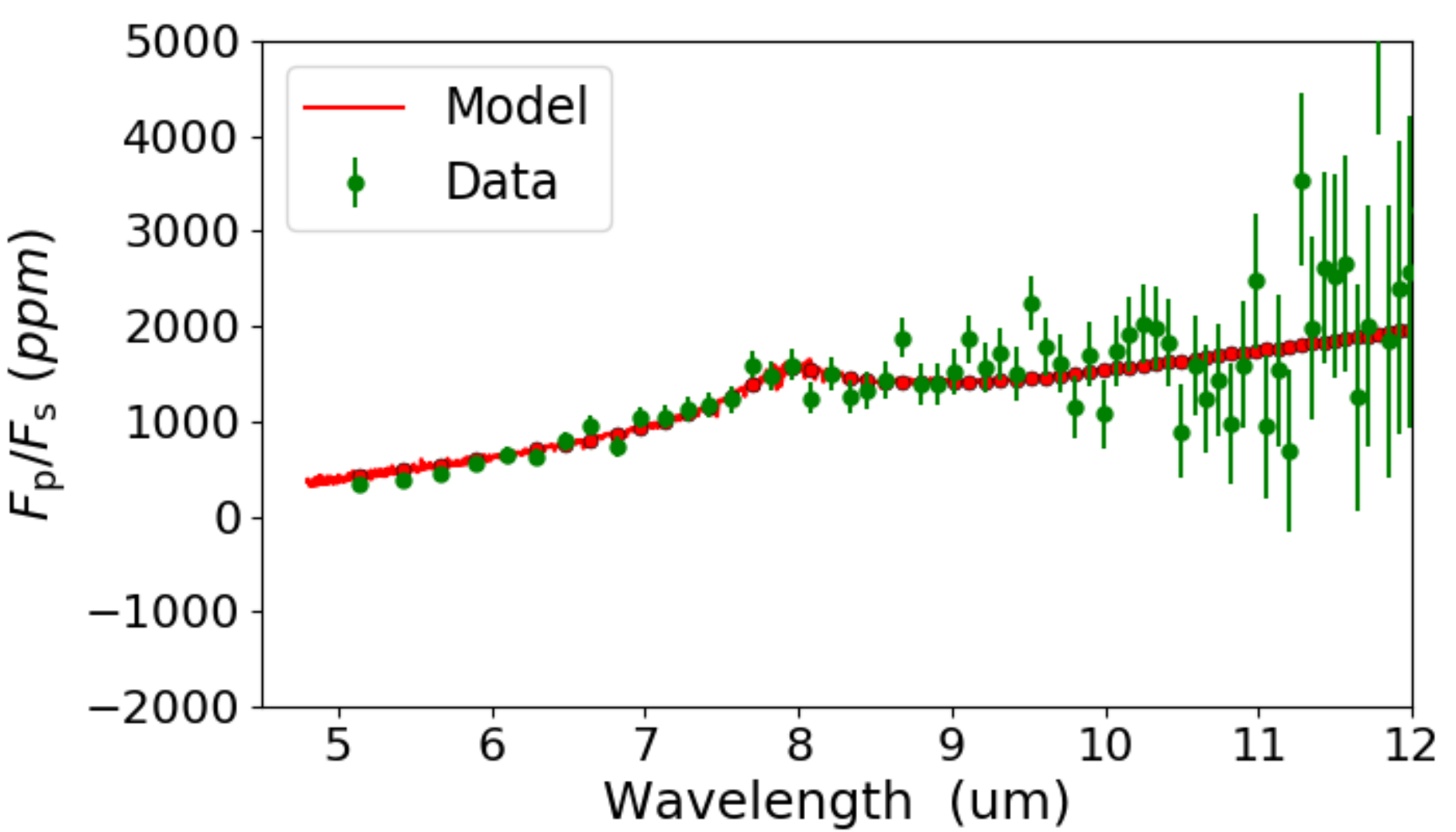}}}
   \vspace{-10pt}
   \subfigure{\includegraphics[height=5.3cm, trim={0.5cm 0 0cm 0}]{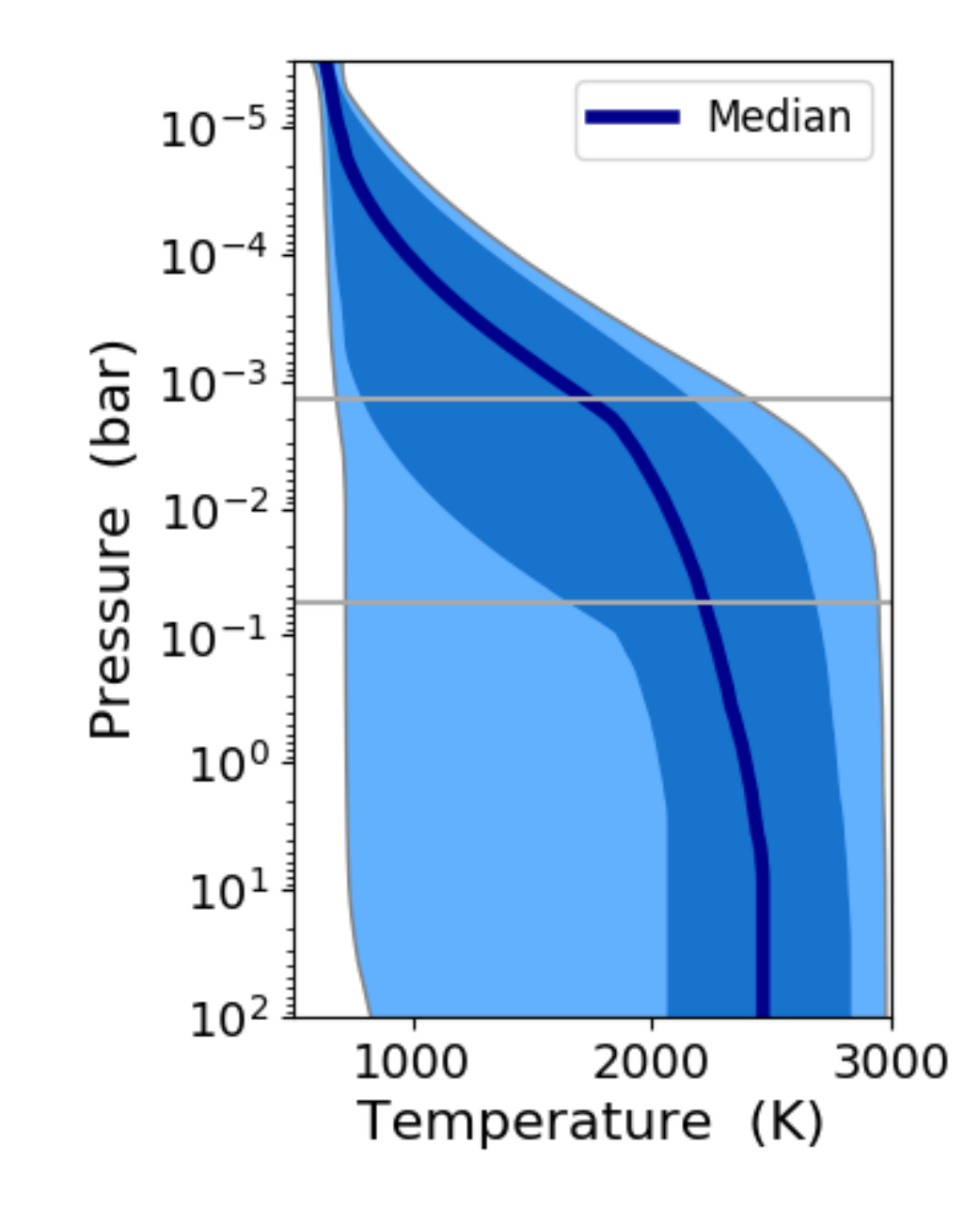}}\hspace{-2pt}
   \vspace{-10pt}
   \subfigure{\includegraphics[height=6.3cm, trim={0.4cm 0 6cm 0}]{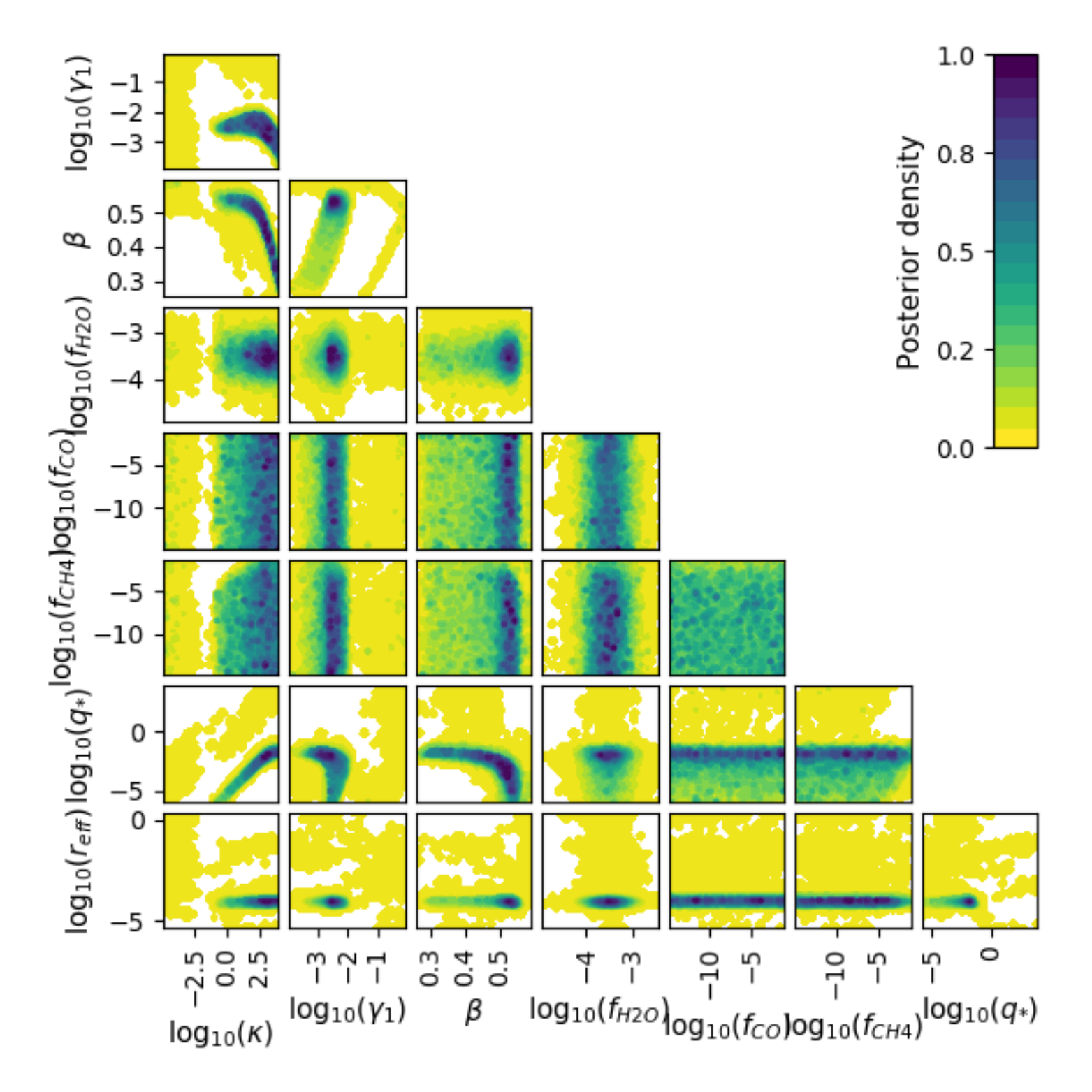}}
   \vspace{-10pt}
   \subfigure{\includegraphics[height=5.5cm, trim={0.5cm 0 0cm 0}]{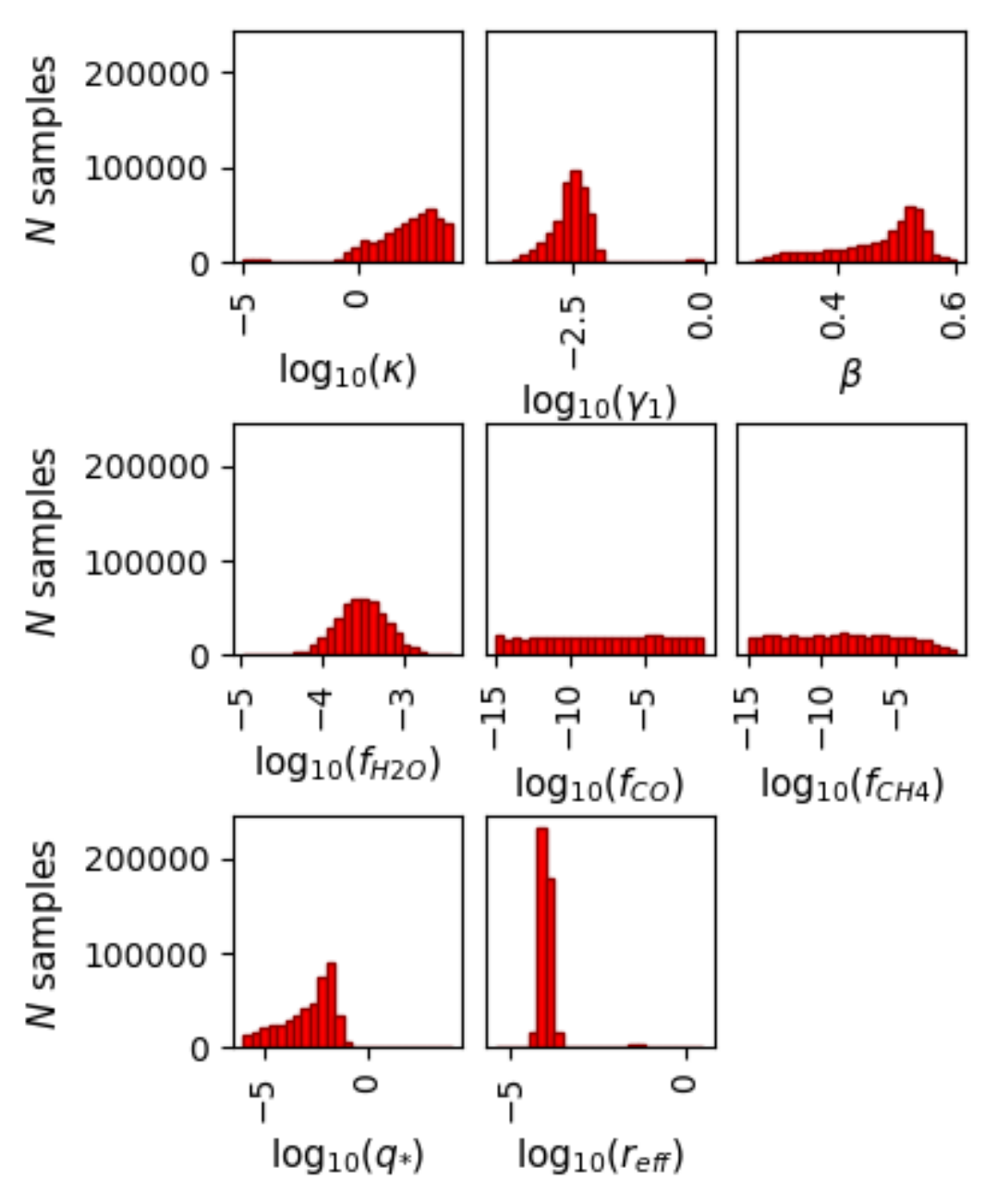}}\hspace{-3pt}
   \subfigure{\raisebox{0.6cm}{\includegraphics[height=4.6cm, clip=True]{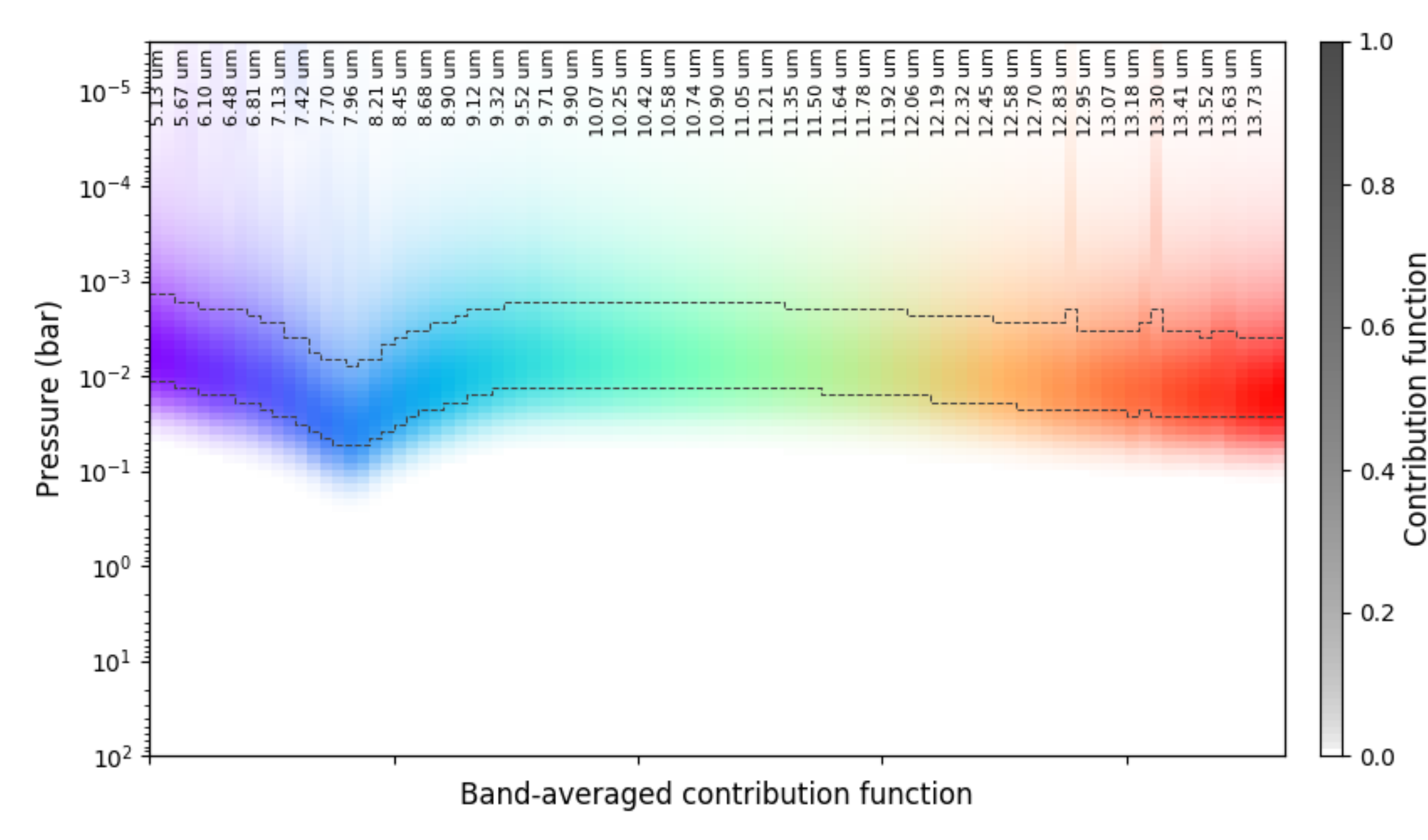}}}
   \caption{$\rm MgSiO_3$ constrained $T-P$ non-equilibrium night-side retrieval. In this run the $T-P$ parameters are allowed to explore only non-inverted profiles, and the species abundances are left to be free. Top panels show the best-fit spectrum model in red with data points and uncertainties in green, the median $T-P$ profile with the extent of the 1 and 2$\sigma$ regions, and in gray the extent of the contribution functions, and on the right the correlations between the parameters with posterior density. Bottom panels show the posterior histograms and the contribution functions.}
   \label{fig:MgSiO3_constrained} 
\end{figure*}

Prior to the {\pyratbay} analysis we ran several different MCMC settings to fully explore the phase space and get the best constraints on the cloud parameters. For all of the tested cases the posterior histograms of some of the cloud parameters were flat with no correlations with other parameters, implying that the data are uninformative about them. Thus, we fixed and excluded them from the exploration ($\Delta i$ was fixed to 80, $H_c$ to 0.75, and $\log_{10}(X_c)$ to 0.0), allowing only the condensate effective particle size, $\log_{10}(r_{\rm eff})$, and the condensate mole fraction, $\log_{10}(q\sp{*})$, to be free. This leaves us with the same parameters as {\taurex}. For the temperature profile in the $\rm MnS$ case, the solution found with {\taurex} is degenerated between $\gamma_1$ and $\gamma_2$. We fix this by disabling the  contribution of the second visible opacity ($\alpha = 0$).

\subsubsection{MnS Clouds}

In the MnS case (Figure \ref{fig:MnS_retrievals}), both {\pyratbay} and {\taurex} fit well the input spectrum model. While the retrieved temperature profiles are hotter than the hemispheric averaged profile, both profiles converged to similar values at the top of the atmosphere where the contribution functions are located. {\water} abundance and cloud mixing ratio were well constrained and closely matching the input value for both codes. As in the cloud-free retrieval, {\pyratbay} used a Gaussian prior on the {\water} abundances ($\log_{10}$(H$_2$O) = -3.52 $\pm$ 0.3), while {\taurex} used an uniform prior (1$\times10^{-1}$--1$\times10^{-5}$). {\pyratbay} retrieved a {\water} abundance to $\log_{10}$(H$_2$O) = -3.48 $\pm$ 0.3, basically the prior value and {\taurex} retrieved $\log_{10}$(H$_2$O) = -4.15$\pm$0.71.  The retrieved cloud particle sizes have a slightly lower value than the input ({\pyratbay} retrieves $\rm log_{10}(r_{\rm eff}) = -0.58 \pm 0.14$ $\mu$m, while {\taurex} retrieves $\rm log_{10}(r_{\rm eff}) = -1.7 \pm 0.33$ $\mu$m). Both codes produced no constraints on the $\rm CO$ abundance, while {\methane} abundances could not be constrained in the {\taurex} run and has produced a lower limit ($\log_{10}$(CH$_4$) > -2.4) in the {\pyratbay} run. The retrieved cloud parameters and calculated chemical abundances of the {\pyratbay} ``self-consistent'' retrieval are nicely matching those of the ``free'' retrieval. A summary of the retrieved abundances is given in Table \ref{tab:comp_MnS}.

Given the relatively similar results both codes obtained in the cloud-free retrieval case (Sect.\ref{sec:cloud-free ret}), differences seen here between the retrieved parameters' values within {\pyratbay} and {\taurex} can be mostly attributed to different cloud parametrization schemes between the two codes: higher complexity of one cloud model compare to the other, different treatment of thermal scattering, larger number of free parameters, more freedom in the shape of the log-normal particle distribution, flexibility in the cloud base location and the shape of the cloud, and possible difference in the resolution of the output models.

\begin{table}[!h]
\caption{Comparison of the retrieved abundances obtained by \\
{\taurex} and {\pyratbay} for the MnS clouds cases. There are no uncertainties for the self-consistent runs, which have been performed with {\pyratbay} only.} \label{tab:comp_MnS}
\centering
\begin{tabular}{c|c|c|c}
\hline
\hline
Case             & Molecule      &  \taurex            &      \pyratbay   \\
\hline
                 & log(H$_2$O)   &  -4.15$\pm$0.71     & -3.48$\pm$0.30\\
Free             & log(CO)       &  < -3               & unconstrained  \\
                 & log(CH$_4$)   &  unconstrained      &  > -2.4   \\
\hline
                & log(H$_2$O)    &   N/A                & -3.1\\
Self-consistent & log(CO)        &   N/A                & -9.5  \\
(0.1 bar)       & log(CH$_4$)    &   N/A                & -3.3   \\
\end{tabular}
\end{table}

\subsubsection{MgSiO$_3$ Clouds}

For the MgSiO$_3$ clouds, we performed the same ``free'' retrieval runs as for the MnS clouds. However, in this case, ``free'' retrievals did not manage to converge to a physically realistic solution for both {\taurex} and {\pyratbay}. The codes appear to misinterpret the bump at around 8 $\mu$m as an emission feature, cut the contribution functions at the very top of the atmosphere, and fit the spectrum with a high {\methane} abundance and an inverted temperature profile (see Figure \ref{fig:MgSiO3_retrievals}). 

To overcome the encountered issue, we guided the temperature-pressure model to explore only non-inverted solutions in both codes. {\taurex} was unable to converge to a physical solution, while {\pyratbay} succeeded to get a match between the retrieved temperature and cloud parameters and the input model. In Figure \ref{fig:MgSiO3_constrained} we show the results of this retrieval. To explore only the non-inverted temperature profiles, in {\pyratbay} we fixed  $\gamma_2$ and $\alpha$ parameters to zero and $\gamma_1$ to values smaller than zero, and let $\kappa$ and $\beta$ parameters to be free. The retrieved condensate particle size is 1 $\mu$m as the input value, $\rm log_{10}(r_{eff}) = -4.069 \pm 0.509$ cm, and the cloud mole fraction upper limit is around $\rm 10^{-2}$. H$_2$O abundances is again closely matching the Gaussian prior, log$_{10}$(H$_2$O) = -3.58 $\pm$ 0.305, and the contribution functions reveal the location of the cloud at $10^{-3}$ bar, at the same level as the input model. The retrieved best-fit spectrum, median temperature profile, and the condensate particle size ($\rm log_{10}(r_{eff}) = -4.039 \pm 0.128$ cm) for the {\pyratbay} ``self-consistent'' retrieval are almost identical to the constrained ``free'' retrieval case, with the cloud mole fraction having similar upper limit, but with the contribution functions located at lower pressures ($\rm 10^{-5}$ bar). The chemical abundances values for all MgSiO$_3$ cases are gathered in Table \ref{tab:comp_MgSiO3}. Apart from the prior value retrieved for the water, neither {\pyratbay} nor {\taurex} are able to provide useful constraints on the chemical abundances. Particularly, in the free retrieval case {\pyratbay} provides a biased measurement of the CH$_4$ abundance. Despite these shortcomings, {\pyratbay} provides a surprisingly good match of the particle size and a realistic upper limit on the cloud height that cannot be obtained with {\taurex}.\\

\begin{table}[!h]
\caption{Comparison of the retrieved abundances obtained by \\
{\taurex} and {\pyratbay} for the MgSiO$_3$ clouds cases.} \label{tab:comp_MgSiO3}
\centering
\begin{tabular}{c|c|c|c}
\hline
\hline
Case            & Molecule      &       \taurex    &      \pyratbay   \\
\hline
                & log(H$_2$O)   & not converged   &  -3.78$\pm$0.29\\
Free            & log(CO)       & not converged   &  < -1.36  \\
                & log(CH$_4$)   & not converged   &  > -3.4   \\
\hline
                & log(H$_2$O)   & not converged   & -3.58$\pm$0.31\\
Constrained     & log(CO)       & not converged   & unconstrained  \\
 T-P            & log(CH$_4$)   & not converged   & unconstrained   \\ 

\hline
                & log(H$_2$O)   &   N/A             & -3.0\\
Self-consistent & log(CO)       &   N/A             & -9.0  \\
(0.1 bar)       & log(CH$_4$)   &   N/A             & -3.3   \\

\end{tabular}
\end{table}

In conclusion, based on the cloudy retrieval analysis, distinguishing between cloud-free and cloudy atmospheres in  \jwst/MIRI data could present a challenge without a careful approach. Even for the 1 $\mu$m synthetic models with $\rm MgSiO_3$ clouds, which show a noticeable silicate feature around 10 $\mu$m, we saw a degenerate solution with high CH$_4$ abundance, inverted temperature profile and no traces of clouds. The challenge will become even higher for the particle sizes larger than 10 $\mu$m, as the silicate feature becomes even less pronounced \citep{Wakeford2015}. 

To overcome this issue, one must closely examine the physical background of the retrieved model (e.g., temperature profile, spectral features seen in emission rather than in absorption, abundances of the retrieved species) and discard non-physical solutions. In addition, the complexity of the underlying parametrized cloud models plays a crucial role. More advanced and realistic cloud models that, for example, include thermal scattering, more freedom in the cloud particle distribution, and the location and shape of the cloud, have higher chances to discard false models and allow Bayesian sampling to land in a realistic phase space. \citet{greene2016} also points out the importance of the synergy between different instruments  on board \jwst to obtain wider wavelength coverage and avoid abundances/cloud properties degeneracies. Combining NIRCam, NIRISS, and MIRI instruments, that cover 0.6-11 $\mu$m range, would be our best way forward \citep{SchlawinEtal2018}. In particular,  \citet{ChuhongLine2019-arXiv} show that NIRCam wavelength range ($\lambda$ = 2.5 - 5 $\mu$m) is critical in inferring atmospheric properties (precise compositional constraints are possible due to the presence of CO and CO$_2$ features at these wavelengths), while NIRISS + MIRI ($\lambda$ = 0.6 - 2.5, 5 - 11 $\mu$m) are necessary for constraints on cloud parameters (NIRISS is required to constrain the scattering slope at shorter wavelengths, while MIRI is important due to the existence of mid-IR resonance features). Finally, a 2.5D approach such as the one of~\citet{Irwin2019} would further improve the ability to retrieve the cloud properties.

Overall, our retrieval models can distinguish between a cloudy and a cloudless nightside, but if the clouds are too thick or have a single-scattering albedo in the infrared that is too large, such as the MgSiO$_3$ case, our state-of-the art retrieval methods have difficulty analysing the complexity of the planned \jwst/MIRI observations. If not carefully guided, they can produce biased detection of a high {\methane} abundance, regardless of a well-known specific absorption feature at 10 $\mu$m. More work is needed to leverage the planned observations and quantify the nightside abundances in the presence of nigthside clouds.

\section{Conclusions}\label{sec:conclusion}

In order to prepare the future observations of WASP-43b that will be carried out during the \jwst Early Release Science Program \citep[PIs: N. Batalha, J. Bean, K. Stevenson;][]{stevenson2016, bean2018}, we performed a series of theoretical models to better understand the atmosphere of this hot Jupiter and predict \jwst/MIRI observations. In addition to predictions of the observed spectra, this work allowed us to compare some outputs obtained with our different models, highlighting uncertainties in atmospheric properties and exploring the robustness and weakness of atmospheric retrievals. The key results of  this study are:

- The thermal structure and the corresponding spectra found with our 2D radiative transfer code (\texttt{2D-ATMO}) was very similar to that found with our more complex Global Circulation Model (\texttt{SPARC/MITgcm}), despite the two codes being based on different physical approaches. Because the 2D code runs much faster than the GCM, it presents a very good alternative for atmospheric studies that need to be rapidly addressed.

- From a chemical aspect, thanks to our 1D and pseudo-2D chemical models of WASP-43b's atmosphere, we found that the main constituents after H$_2$/He were CO, H$_2$O and N$_2$. Methane is not expected to be abundant in the layers probed by observations, but HCN, photochemically produced on the dayside, could be the second-most important C-bearing species after CO in these regions.

- Thanks to our microphysical clouds model, we determine that the night side of WASP-43b is probably cloudy (MnS, Na$_2$S, Mg$_2$SiO$_4$/MgSiO$_3$). The cloud coverage of the dayside depends on whether silicates form a deep cloud. This depends on microphysics (e.g. forsterite vs. enstatite formation), atmospheric dynamics (strenght of vertical mixing) and the deep energetics (how hot is the planet interior).

- By simulating the observations and performing a data retrieval, we showed that this full orbit spectroscopic phase curve should allow us to constrain the nightside pressure-temperature structure, and the {\water}, {\methane}, and CO abundances if the atmosphere is not cloudy. We would be able to conclude whether horizontal quenching drives the nightside atmosphere out of chemical equilibrium. If the atmosphere is cloudy we should be able to detect the presence of clouds, differentiate between different cloud chemical composition and constrain the mean particle size with retrieval models that include mie scattering such as {\pyratbay}. However, quantifying the nightside atmospheric composition, thermal structure, and cloud abundance will be more challenging and requires more careful preparation.

\acknowledgments
This work was supported by the Centre National d'\'Etudes Spatiales (CNES). The research leading to these results has received funding from the European Union's Horizon 2020 Research and Innovation Programme, under Grant Agreement No 776403, grant agreement No 758892 (ExoAI) and grant agreement No 757858 (ATMO). This project has received also support from NASA through a grant from STScI (JWST-ERS-01366). O.V. thanks the CNRS/INSU Programme National de Plan\'etologie (PNP) for funding support. P.-O.L. and P.T. thanks the LabEx P2IO. I.P.W. and C.Q. acknowledge funding by the Science and Technology Funding Council (STFC) grants: ST/K502406/1, ST/P000282/1, ST/P002153/1 and ST/S002634/1. J.M. thanks the NASA Exoplanet Research Program grant number NNX16AC64G. J.B. and I.D.D. thank to the NASA Exoplanet Research Program grant number NNX17AC03G. M.S. was supported by NASA Headquarters under the NASA Earth and Space Science Fellowship Program - Grant 80NSSC18K1248. S.C. was supported by an STFC Ernest Rutherford Fellowship. We thank Jake Taylor for comments on a draft version of the manuscript. We thank the referee for his careful reading of the manuscript.

\bibliographystyle{yahapj}
\bibliography{references}


\end{document}